\documentclass[pra,a4paper,twoside,twocolumn,secnumarabic,balancelastpage,amsmath,amssymb,hyperref=pdftex,superscriptaddress 
]{revtex4}

\usepackage{graphics}      
\usepackage[pdftex]{graphicx}      
\usepackage{longtable}     
\usepackage{bm}            
\usepackage{verbatim}
\usepackage{mathtools}
\usepackage{adjustbox}
\usepackage[colorlinks=true]{hyperref}

\begin{document}

\title{Collective mode in the SU(2) theory of cuprates}

\date{\today}

\author{C. Morice}
\affiliation{Institut de physique th\'eorique, Université Paris Saclay, CEA, CNRS, F-91191 Gif-sur-Yvette, France}
\author{D. Chakraborty}
\affiliation{Institut de physique th\'eorique, Université Paris Saclay, CEA, CNRS, F-91191 Gif-sur-Yvette, France}
\author{C. P\'epin}
\affiliation{Institut de physique th\'eorique, Université Paris Saclay, CEA, CNRS, F-91191 Gif-sur-Yvette, France}

\begin{abstract}
Recent advances in momentum-resolved electron energy-loss spectroscopy (MEELS) and resonant inelastic X-ray scattering (RIXS) now allow one to access the charge response function with unprecedented versatility and accuracy. This allows for the study of excitations which were inaccessible recently, such as low-energy and finite momentum collective modes. The SU(2) theory of the cuprates is based on a composite order parameter with SU(2) symmetry fluctuating between superconductivity and charge order. The phase where it fluctuates is a candidate for the pseudogap phase of the cuprates. This theory has a signature, enabling its strict experimental test, which is the fluctuation between these two orders, corresponding to a charge 2 spin 0 mode at the charge ordering wave-vector. Here we derive the influence of this SU(2) collective mode on the charge susceptibility in both strong and weak coupling limits, and discuss its relation to MEELS, RIXS and Raman experiments. We find two peaks in the charge susceptibility at finite energy, whose middle is the charge ordering wave-vector, and discuss their evolution in the phase diagram.
\end{abstract}

\maketitle

\section{Introduction}

Cuprate superconductors exhibit a very rich phenomenology. They display an antiferromagnetic phase at low doping, and a $d$-wave superconducting phase at higher doping. The pseudogap phase, detected above this superconducting dome, corresponds to a partial gapping of the Fermi surface in the parts of the Brillouin zone furthest away from the nodes of the $d$-wave gap, named antinodal regions \cite{Alloul89, Alloul91, Warren89, Campuzano98, Vishik:2012cc, Vishik12, Yoshida:2012kh, He14, Vishik14}. In the same doping region, charge order was observed under applied magnetic field by quantum oscillations and transport measurements \cite{Doiron-Leyraud07, LeBoeuf07, LeBoeuf11, Laliberte11, Sebastian12, Doiron-Leyraud13, Barisic2013, Grissonnanche:2015tl}. X-ray scattering exposed incipient charge modulations with incommensurate wave vectors at zero field \cite{Ghiringhelli12, Chang12, Achkar12, Blanco-Canosa13, Blackburn13a, Blackburn13b, Thampy13, Blanco-Canosa14, Tabis14, Comin14, Comin14a, Comin:2015vc, Comin:2015ca, Forgan2015, Chang16}.

The observation of the charge order led to the development of theories exploring a connection between superconductivity and charge order as the origin of the pseudogap. In particular, in the SU(2) theory, a composite order parameter describes both superconductivity and charge order \cite{Efetov13, Kloss15, Kloss:2016hu, Montiel16, Montiel2017sr}. The pseudogap phase is then the phase in which the composite order parameter has a finite length but fluctuates between the two states. It was shown to agree with a range of experimental responses, including ARPES \cite{Montiel:2016it}, Raman scattering \cite{Montiel15a}, inelastic neutron scattering \cite{MontielNeutrons2017}, transport measurements \cite{Morice2017transport} and high magnetic field studies \cite{Chakraborty2018}.

Out of the large variety of models attempting to describe the pseudogap, many theories have tried to explain the partial gapping of the Fermi surface by the presence of an antiferromagnetic mode \cite{Abanov00, Tchernyshyov2001, Abanov03, Hao2009, Metlitski10a, Metlitski10b}, or the rotation of the superconducting order parameter to another channel, such as a staggered-flux phase \cite{Lee06}. In the SO(5) theory, there is a single order parameter whose coordinates correspond to antiferromagnetism and superconductivity, and the pseudogap phase corresponds to the phase in which this order parameter fluctuates between the two \cite{Demler95, Demler1997, Zhang1997, Zaleski2000, Zaleski2001, Hu2001, Demler2004}. One could however also think that the antiferromagnetic phase could be detached from the pseudogap, and both these phases could be different types of condensation at the antiferromagnetic coupling energy scale $J$.

In order to check whether the SU(2) theory adequately describes the pseudogap phase in the cuprates, we need to find its unique signature which could be used as an experimental test of its validity. Because it features a composite order parameter which fluctuates, the SU(2) theory has a very specific signature in the form of a collective mode. Indeed, if a theory is based on the fluctuations of a composite order parameter, the collective mode corresponding to these fluctuations can often be probed directly. Other works have also explored the connection between superconducting and charge order parameters, often in the context of competing orders \cite{Kivelson:2002er, Zhang:2002hz}. These formalisms do not feature an enhanced symmetry and therefore do not exhibit the kind of collective modes discussed here. However some other works have considered composite order parameters such as the one we study here, and our work is therefore potentially relevant to these \cite{Hayward14, PhysRevB.92.224504, PhysRevLett.119.107002}.

This collective mode, in the case of the SU(2) theory, bears charge 2 and spin 0, and thus has to be studied in the charge channel. Moreover it peaks at the ordering wave-vector of the charge density wave. Until very recently, a highly-sensitive fully momentum-resolved charge probe did not exist, and therefore there was no hope to study the signature mode of the SU(2) theory. Recently, a new experimental technique named momentum-resolved electron energy-loss spectroscopy (MEELS) was developed \cite{Vig2017, Mitrano2017}. It allows one to probe the charge response resolved both in momentum and frequency spaces, with very high resolution. Interestingly for the study of cuprates, it can probe both normal and superconducting states interchangeably. Its versatility has led to its comparison with ARPES in probing directly fundamental degrees of freedoms \cite{Vig2017}.

Resonant inelastic X-ray scattering (RIXS) has recently also been improved to the point where energy resolution has reached 40 meV, making it a great tool for the study of charge excitations at finite momentum and low energy. Raman scattering has been used extensively to probe the charge response in cuprate superconductors. Finally, optical conductivity has been recently shown to be an interesting tool to study collective modes in superconductors \cite{Moor2017}.

Here we describe the collective mode in the SU(2) theory of cuprates. This mode bears charge two and spin zero, is centred on the charge ordering wave vector, and corresponds to a pair density wave (PDW). It has a finite resonance energy, and disperses away from it; we give a theoretical estimate of the slope and find that it fits experiments. Its influence on the imaginary part of the charge susceptibility is limited to the superconducting phase, inside the Stoner continuum region. It peaks at the two crossing points of the dispersion of the collective mode and of the Stoner continuum, close to the charge density wave ordering wave-vector, at a frequency close to twice the superconducting gap. These two peaks in the charge susceptibility go away from each other when applying a magnetic field. We start by describing the system in the strong coupling limit by enforcing an SU(2) constraint on the charge and superconducting orders, and derive a non-linear $\sigma$-model describing the system. We then turn to the weak coupling limit and derive the mode using a self-consistent linear response formalism. Next we discuss how MEELS, RIXS and Raman experiments can access this resonance. Finally, in the two last sections, we calculate the contribution of the mode to the charge susceptibility in the strong and weak coupling formalisms.

\section{Strong coupling between charge and superconducting order parameters}
\label{Strong coupling regime}

In this section, we describe the strong coupling regime of the SU(2) theory of the pseudogap, meaning that we study a theory where the charge order parameter and superconductivity are infinitely coupled. The starting point is to set an order parameter, displayed on Fig.\ \ref{fig:order-parameter}A, which can either condense as a superconductor or a charge order, or fluctuate between these two orders. The strong coupling between these two orders is then obtained by setting a condition on the magnitude of the order parameter, effectively setting a relationship between the magnitudes of the two orders. This allows us to describe the fluctuation of our order parameter in terms of a non-linear $\sigma$-model which describes the collective modes corresponding to the angular fluctuations of the order parameter.

\subsection{Order parameter}

In BCS superconductors, the energy scale corresponding to the formation of Cooper pairs, i.e.\ the pairing potential, is associated with the transition temperature $T_c$. In cuprate superconductors, one important energy scale is the antiferromagnetic coupling constant $J$. It is of order 1500 K \cite{Lee06}, much larger than the observed $T_c$. This means that in this case there are at least two energy scales in the problem: $J$ and $T_c$ \cite{Zhang1997}. These two energies can be associated respectively with the one at which the magnitude of the composite order parameter chosen for the pseudogap becomes finite and with the one at which its phase becomes fixed. This means that at energies between these two the system is fluctuating: it exhibits a composite order parameter with a finite magnitude but with a fluctuating phase.

This does coincide with experimental observations in cuprate compounds. At intermediate doping, the system features a high-energy phase, namely the pseudogap phase. It also features two low-energy phases: the superconducting phase and the charge ordered phase at high applied magnetic field. One can therefore, in the framework of the SU(2) theory, associate the high energy scale $J$ with the forming of either particle-particle or particle hole pairs. The pseudogap phase then corresponds to fluctuations between two different types of condensation: superconductivity for particle-particle pairs and charge order for particle-hole pairs, which happen at very similar temperatures.

In order to encompass these various possibilities, one can use an effective model that takes them simultaneously into account. At sufficiently high doping, one can consider neglecting the antiferromagnetic order, since it only emerges at low doping. We are therefore left with charge and superconducting orders, whose relation has been studied intensively in recent years \cite{Efetov13}. A simple model embracing these two orders is one where we define a composite order parameter describing both orders.
We first need to decide what charge order to consider. We adopt a very general perspective and only limit ourselves to orders which will enable us to define a closed algebra. We define a involution $\textbf{k} \rightarrow \overline{\textbf{k}}$, i.e.\ a function which is its own inverse and maps the Brillouin zone on itself, such as:
\begin{equation}
\overline{\overline{\textbf{k}}} = \textbf{k}
,
\overline{(-\textbf{k})} = -\overline{\textbf{k}}
\label{eq:involution}
\end{equation}
which we use to define a simple $d$-wave charge order parameter:
\begin{equation}
\chi = \frac{1}{2} \sum_{\textbf{k} \sigma} \overline{d}_\textbf{k} c^\dagger_{\overline{\textbf{k}} \sigma} c_{-\textbf{k} \sigma}
\end{equation}
where $\overline{d}_\textbf{k} = \cos(2\theta_\textbf{k}) + \cos(2\theta_{\overline{\textbf{k}}})$ is a $d$-wave factor, $\theta_\textbf{k}$ is the argument of $\textbf{k}$ and $\sigma$ is a spin index. Note that the $d$-wave factor is invariant both by the involution and by space inversion $\textbf{k} \rightarrow -\textbf{k}$. The charge density wave is the main observed instability in the under-doped phase of the cuprates, hence here we choose to rotate from superconductivity to this frame. But we could as well have enlarged the rotation space and chosen to rotate as well to incommensurate currents. This would not change the structure of the collective modes but add some other channels for collective modes.

\begin{figure}
\centering
\includegraphics[width=8cm]{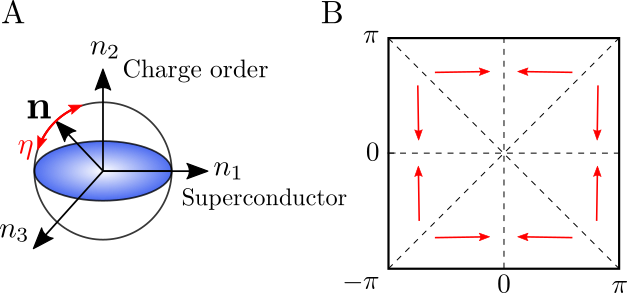}
\caption{(A) We describe the system by an order parameter $\textbf{n}$ which has three components: two associated with superconductivity and one with charge order. The pseudogap phase is associated with the phase where this order parameter becomes finite in length and its phase fluctuates. The fluctuation between the charge order component $n_2$ and the superconducting components $n_1$ and $n_3$ is the $\eta$ mode, corresponding to a PDW, depicted in red. (B) The charge order parameter $\textbf{Q}_0$, depicted in red, is $\textbf{k}$-dependent. Here we choose to divide the Brillouin zone in eight quadrants, following \cite{Montiel2017}.}
\label{fig:order-parameter} 
\end{figure}

The involution used to define the charge order parameter can be taken to be $\overline{\textbf{k}} = -\textbf{k} + \textbf{Q}_0$, where the charge ordering wave vector $\textbf{Q}_0$ connects two hot-spots, which are the points of the Fermi surface where the antiferromagnetic fluctuations diverge. This expression for $\overline{\textbf{k}}$ can only be involutive, i.e.\ verify the first equation in Eq.\ \eqref{eq:involution}, if $\textbf{Q}_0$ is $\textbf{k}$-dependent. In the following we will use a general formalism everywhere possible, but when necessary we will set $\textbf{Q}_0 = \left( \pm Q_x, 0 \right), \left( 0, \pm Q_y \right)$ where $Q_x$ and $Q_y$ are the distance between two hot-spots for the electronic dispersion we will be considering. The $\textbf{k}$-dependence is set to be per quadrant as displayed on Fig.\ \ref{fig:order-parameter}B, following \cite{Montiel2017}. Note that other types of involutions, such as one that maps one opposite sides of the Fermi surface on each other, have also been considered in previous works \cite{Montiel2017}.

Following experimental observations \cite{Vishik2012} we consider a $d$-wave superconducting order parameter:
\begin{equation}
\Delta = \frac{1}{\sqrt{2}} \sum_{\textbf{k}} \overline{d}_\textbf{k} c_{\textbf{k} \downarrow} c_{-\textbf{k} \uparrow}
\end{equation}
which we can now use to define a composite order parameter. The choice that has been made in \cite{Efetov13} is to use the matrix:
\begin{equation}
\begin{pmatrix}
\chi & \sqrt{2} \Delta\\
- \sqrt{2} \Delta^\dagger & \chi
\end{pmatrix}
\text{  with  }
2 |\Delta|^2 + |\chi|^2 = 1
\label{eq:su2-condition}
\end{equation}
which belongs to the SU(2) group. Here, for convenience, we set the value of the constraint in the right-hand side of Eq.\ \eqref{eq:su2-condition} to be 1. But more generally, it can take any value and it sets the energy scale of the pseudogap.

One could also consider other types of low-temperature orders to which associate a component of such a composite order parameter. In particular, at low doping, cuprate superconductors feature an antiferromagnetic phase.

We now want to describe the symmetry of this composite order parameter, and in order to do this we want to write a Lie algebra that will act on our order parameter space. It is natural, in this case, to choose su(2) as the Lie algebra which will act on our SU(2) order parameter (by convention, we write groups in capital letters and Lie algebras in lower-case). It is also interesting to note that SU(2) is closely related to the SO(3) group, more particularly by a mapping which maps one element of SO(3) on two elements of SU(2), i.e.\ a covering map of order 2. However, Lie algebras are local objects, and because of this, one can always place themselves in a neighbourhood of SU(2) where the mapping will be of order one, hence su(2) and so(3) are isomorphic.

One can therefore define an order parameter on which to act with an so(3) Lie algebra, which will be simpler to handle and closer to relate with other phenomenological theories \cite{Zhang1997}. so(3), being isomorphic to su(2), will then satisfactorily preserve the symmetry of the order parameter, i.e.\ preserve the condition in Eq.\ \eqref{eq:su2-condition}. We define the Hermitian order parameter $\textbf{n}$:
\begin{align}
n_1 &= \frac{1}{\sqrt{2}} \left( \Delta^\dagger + \Delta \right)\\
n_2 &= \chi\\
n_3 &= -\frac{i}{\sqrt{2}} \left( \Delta^\dagger - \Delta \right)
\end{align}
The state of the system is therefore described by a three-dimensional vector (Fig.\ \ref{fig:order-parameter}A). Two of its components are associated with superconductivity, while the third one is associated with charge order. Thus, in the charge ordered phase, $\textbf{n}$ is along $n_2$, in the superconducting phase, $\textbf{n}$ is in the $n_1$-$n_3$ plane. In the pseudogap phase, the length of $\textbf{n}$ is finite, and its phase fluctuates between these axes. At temperatures above the pseudogap transition temperature $T^*$, $\textbf{n}$ vanishes.

\subsection{Lie algebra}

Now that we have defined the order parameter we can define a Lie algebra to relate its coordinates to one another. The Lie algebra will act on the order parameter space via its adjoint representation, meaning that an element $x$ of the Lie algebra will transform an element $v$ of the order parameter vector space following:
\begin{equation}
v \rightarrow \left[ x,v \right]
\end{equation}
This defines the action of the Lie algebra on the order parameter vector space, which is an endomorphism of the order parameter space. Let us now relate the three coordinates of $\textbf{n}$ to one another using such commutators. The operators we will use for this will then be the generators of our Lie algebra, and will thus define it entirely.

We define, following previous work \cite{Montiel2017}:
\begin{equation}
\eta^\dagger = \sum_\textbf{k} c^\dagger_{\textbf{k} \uparrow} c^\dagger_{\overline{\textbf{k}} \downarrow}
\label{eq:eta-operator}
\end{equation}
\begin{equation}
\eta_z = \left[ \eta^\dagger , \eta \right] = \frac{1}{2} \sum_{\textbf{k}} \left( c^\dagger_{\textbf{k} \uparrow} c_{\textbf{k} \uparrow} + c^\dagger_{\textbf{k} \downarrow} c_{\textbf{k} \downarrow} - 1 \right)
\end{equation}
which, if we use a $l=1$ triplet notation $\Delta_{-1} = \Delta$, $\Delta_{0} = \chi$, $\Delta_{1} = \Delta^\dagger$, $\eta^\dagger = \eta^+$, and $\eta = \eta^-$ verify the relations:
\begin{equation}
\left[ \eta^\pm , \Delta_m \right] =  \sqrt{l (l+1) - m (m \pm 1)} \Delta_{m \pm 1}
\end{equation}
\begin{equation}
\left[ \eta_z , \Delta_m \right] =  m \Delta_m
\end{equation}

The two superconducting components $n_1$ and $n_3$ are related via a U(1) symmetry, whose generator is the total charge:
\begin{equation}
Q = \eta_z + \frac{1}{2} = \frac{1}{2} \sum_{\textbf{k} \sigma} c^\dagger_{\textbf{k} \sigma} c_{\textbf{k} \sigma}
\end{equation}
Indeed we can verify the relationship:
\begin{equation}
\left[ Q, \Delta \right] = \frac{1}{2\sqrt{2}} \sum_{\textbf{k}_1 \textbf{k}_2 \sigma} \overline{d}_{\textbf{k}_2} \left[ c^\dagger_{\textbf{k}_1 \sigma} c_{\textbf{k}_1 \sigma}, c_{\textbf{k}_2 \downarrow} c_{-\textbf{k}_2 \uparrow} \right] = - \Delta
\end{equation}
which gives us:
\begin{equation}
\left[ Q, n_1 \right] = \frac{\left[ Q, \Delta^\dagger \right] + \left[ Q, \Delta \right]}{\sqrt{2}} = -\frac{\left[ Q, \Delta \right]^\dagger + \left[ Q, \Delta \right]}{\sqrt{2}} = i n_3
\end{equation}
which relates $n_1$ and $n_3$, the two superconducting components of $\textbf{n}$, via the action of the total charge $Q$.

The $\eta^\dagger$ operator corresponds to an s-wave pair density wave: it pairs two electrons of opposite spin at a finite momentum $\textbf{Q}_0$. Here it corresponds to the rotation from the superconducting to the charge order parameter, unlike in recent works where it is at the origin of the formation of the charge density wave order \cite{Dai2018}.

Combining $\eta^\dagger$ with $\eta$ allows us to relate $n_1$ to $n_2$:
\begin{equation}
\left[ i\frac{\eta^\dagger - \eta}{2} , n_1 \right] = i n_2
\end{equation}
\begin{equation}
\left[ -i\frac{\eta^\dagger - \eta}{2} , n_2 \right] = i n_1
\end{equation}
and similarly for $n_2$ and $n_3$:
\begin{equation}
\left[ \frac{\eta^\dagger + \eta}{2} , n_3 \right] =  i n_2
\end{equation}
\begin{equation}
\left[ -\frac{\eta^\dagger + \eta}{2} , n_2 \right]  = i n_3
\end{equation}
We can now gather the generators of our Lie algebra in a single matrix $\textbf{L}$:
\begin{equation}
\textbf{L} =
\begin{pmatrix}
0 & i\frac{\eta^\dagger - \eta}{2} & Q\\
-i\frac{\eta^\dagger - \eta}{2} & 0 & -\frac{\eta^\dagger + \eta}{2}\\
-Q & \frac{\eta^\dagger + \eta}{2} & 0
\end{pmatrix}
\end{equation}
which acts on $\textbf{n}$ via the relation:
\begin{equation}
\left[ L_{ab}, n_c \right] = i\delta_{ac} n_b - i\delta_{bc} n_a
\end{equation}
Note that it satisfies the commutation relation:
\begin{equation}
\left[ L_{ab}, L_{cd} \right] = i \delta_{ac} L_{bd} + i \delta_{bd} L_{ac} - i \delta_{ad} L_{bc} - i \delta_{bc} L_{ad}
\end{equation}

Note that, unlike in other similar theories, the commutation relation between elements of the Lie algebra and of the vector space is strictly verified, and one does not need to take a particular limit such as the continuum limit for the SO(5) theory \cite{Zhang1997}. Indeed, in the SO(5) case, the antiferromagnetic order parameter is related to the $d$-wave superconducting order parameter. One therefore needs to take the continuum limit to take away this $d$-wave form factor. Here, both orders are $d$-wave, hence there is no need to take such a limit.

\subsection{Non-linear $\sigma$-model}

Now that we have defined the composite order parameter and the Lie algebra that relates its components to one another, we can set up the model based on this order parameter. The central characteristic of this model is that in the pseudogap phase the order parameter has a fixed length and its phase fluctuates. We therefore can think of the high energy scale $J$ as corresponding to a mean-field transition below which the length of $\textbf{n}$ becomes finite. Below this energy scale, we neglect changes in the magnitude of the order parameter and simply take $|\textbf{n}|=1$. In order to describe these fluctuations, it is natural to adopt a model where the kinetic energy is that of a rotor (Fig.\ \ref{fig:order-parameter}A), and where the potential energy is an expansion in the gradient of $\textbf{n}$. Moreover, we want this model to describe how the order parameter becomes either superconducting or charge ordered at low temperature. This is done by adding a potential term, small compared to $J$, which can vary with tuning parameters such as the applied magnetic field $B$. This potential term corresponds to the low energy scale associated with $T_c$ at low $B$ and the charge order transition temperature at high $B$.

We take the kinetic energy to be the one of a rotor: $\sum_{a<b} \frac{1}{2 \chi} L_{ab}^2 (x)$ where $\chi$ is the moment of inertia of the rotor. In the long-wavelength limit, one can expand the gradient dependence of the Hamiltonian and obtain: $\sum_{a<b} \frac{\rho_s}{2} v_{ab}^2(x)$ where $v_{ab}^k = n_a \nabla n_b - n_b \nabla n_a$ is a generalised gradient term. These considerations lead us to study the following Hamiltonian density:
\begin{equation}
H = \sum_{a<b} \frac{1}{2 \chi} L_{ab}^2 (x) + \sum_{a<b} \frac{\rho_s}{2} v_{ab}^2(x) + V(\textbf{n})
\label{eq:hamiltonian}
\end{equation}
where $V$ is the potential term.
We want to obtain the corresponding Lagrangian density $\mathcal{L}$ by performing a Legendre transformation \cite{Zhang1997}. We start by expressing the components of $\textbf{L}$ as \cite{Sachdev98}:
\begin{equation}
L_{ab} = n_a p_b - n_b p_a
\label{eq:L}
\end{equation}
where $\textbf{p}$ denotes the conjugated momentum of $\textbf{n}$.
The Legendre transformation can be written as:
\begin{equation}
\mathcal{L} = \dot{\textbf{n}} \textbf{p} - H
\label{eq:legendre-transformation}
\end{equation}
where
\begin{equation}
\dot{\textbf{n}} \textbf{p} = \sum_a \dot{n}_a p_a
\label{eq:legendre-definition}
\end{equation}
and where
\begin{equation}
\dot{n}_c = \frac{\partial H}{\partial p_c}
\end{equation}
We replace $H$ in this latter equation with Eq.\ \eqref{eq:hamiltonian} in which we input Eq.\ \eqref{eq:L} and obtain:
\begin{equation}
\dot{n}_c = \sum_{a \neq c} \frac{1}{\chi} n_a (n_a p_c - n_c p_a)
\label{eq:n-dot}
\end{equation}
We use this expression to rewrite the first term of the Legendre transformation in Eq.\ \eqref{eq:legendre-transformation}, defined in Eq.\ \eqref{eq:legendre-definition}:
\begin{equation}
\dot{\textbf{n}} \textbf{p} = \sum_{a < b} \frac{1}{\chi}(n_a p_b - n_b p_a)^2
\label{eq:legendre-term}
\end{equation}
We define the angular velocity of the rotor:
\begin{equation}
\omega_{ab} = n_a \dot{n}_b - n_b \dot{n}_a
\end{equation}
in which we insert Eq.\ \eqref{eq:n-dot} to obtain:
\begin{equation}
\omega_{ab} = \frac{1}{\chi} \left( n_a p_b - n_b p_a \right)
\end{equation}
We use this expression in Eq.\ \eqref{eq:legendre-term}
\begin{equation}
\dot{\textbf{n}} \textbf{p} = \sum_{a < b} \chi (\omega_{ab})^2
\end{equation}
we use this latter expression in Eq.\ \eqref{eq:legendre-transformation} to finally obtain:
\begin{equation}
\mathcal{L} = \sum_{a<b} \frac{\chi}{2} \omega_{ab}^2 (x) - \sum_{a<b} \frac{\rho_s}{2} v_{ab}^2(x) - V(\textbf{n})
\label{eq:lagrangian}
\end{equation}
We enforce the condition $|\textbf{n}|=1$ by writing the partition function of the system as:
\begin{equation}
\mathcal{Z} = \int D[\textbf{n}] \delta(\textbf{n}^2 - 1) e^{-\int_0^\beta d\tau \int d^x \mathcal{L}}
\end{equation}
The potential term $V$ has to be the same for $n_1$ and $n_3$ in order to preserve the U(1) symmetry of the superconducting state. However it can be different for $n_2$:
\begin{equation}
V(\textbf{n}) = m_{SC} (n_1^2 + n_3^2) + m_{CO} n_2^2
\end{equation}
where $m_{SC}$ is the mass corresponding to the superconducting coordinates and $m_{CO}$ is the mass corresponding to the charge order coordinate. In this case, at low temperature, the system will prefer the order with the larger negative mass. Without an applied magnetic field, the system is known experimentally to be superconducting at low temperature, so we choose the mass of superconductivity to be the larger negative one. Therefore the system will select the corresponding ``easy plane'', i.e.\ at low temperature it will lay in the $n_1$-$n_3$ plane. The mass of the superconducting coordinates is renormalised by applied magnetic field \cite{Chakraborty2018}, which eventually will make it equal to the mass of the charge order coordinate. Beyond this point, the system will prefer charge order at low temperature: there is a spin-flop transition from the superconducting ``easy plane'' to charge order. This type of transition is very flat and corresponds very well to the sharp onset of charge order at high magnetic field in the cuprates \cite{Chakraborty2018}. Note that if $V$ goes beyond second order in the coordinates, the system can also form a phase where both orders coexist \cite{Chakraborty2018}.

\begin{figure}
\centering
\includegraphics[width=8cm]{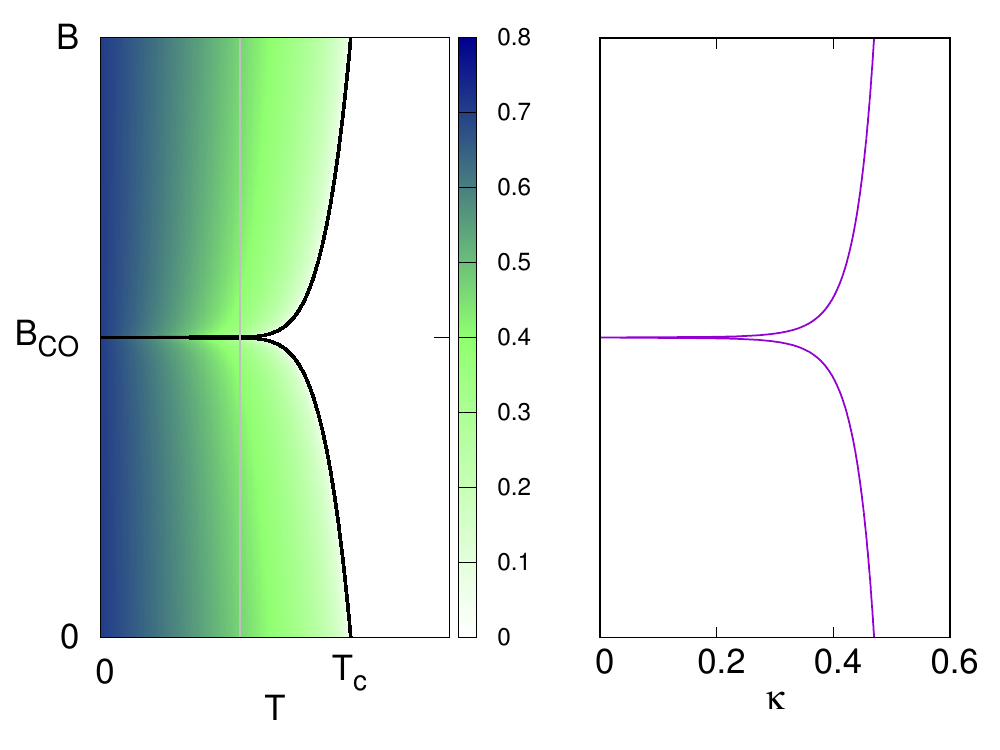}
\caption{Left: evolution of the mass of the $\eta$ mode $\kappa$ with temperature ($T$) and applied magnetic field ($B$) in the superconducting and charge order phases. $B_{CO}$ is the onset magnetic field for the charge ordered phase, and $T_c$ is the superconducting temperature at zero field. Right: evolution of $\kappa$ with magnetic field at the temperature corresponding to the dashed grey line in the left hand side plot. The mass softens at the transition between the two orders. This softening very specifically corresponds to the nature of the $\eta$ mode which rotates from superconductivity to charge order. As such, it is a signature of the SU(2) order parameter. In the case where the two phases coexist, the $\eta$ mode, which corresponds to a PDW, condenses and forms an ordered PDW state showing supersolidity \cite{Chakraborty2018}.}
\label{fig:mass} 
\end{figure}

Thanks to distinguishing between high and low energy scales, we can therefore give a full picture of the phase diagram of the cuprates in the magnetic field ($B$) - temperature ($T$) plane. Antiferromagnetic correlations in the system, corresponding to the high energy scale $J$, cause either particle-particle or particle-hole pairing which we describe with a fixed-magnitude order parameter. Below the higher energy scale but above the lower one, the pairs ``hesitate'' between a zero-charge but finite momentum pairing (charge order) and a $2e$-charge zero-momentum pairing (superconductivity), i.e.\ the phase of this composite order parameter is left free and fluctuates between the two orders. Below the lower energy scale, depending on the magnetic field, the system then chooses one state, where the phase does not fluctuate anymore.

The fact that this phase can fluctuate means that there are collective modes, which correspond to all the ways these fluctuations can take place.

\subsection{Collective modes}
\label{Collective mode}

Now that we have written the non-linear $\sigma$-model for our order parameter $\textbf{n}$ (Eq.\ \eqref{eq:lagrangian}) we can study how it fluctuates. Each type of fluctuation will correspond to a bosonic collective mode. In a BCS superconductor, there is only one mode associated with the phase fluctuation of the superconducting order parameter, whose energy is pushed up to the plasma frequency by a long-range Coulomb interaction \cite{Anderson:1963vi}. In our case the order parameter has another coordinate, which means that there will also be a collective mode associated with the rotation between the superconducting coordinates and the charge order coordinate. We will call this mode the $\eta$ mode, in analogy with the name of the ladder operator relating these two orders (Eq.\ \ref{eq:eta-operator}). Since the $\eta$ operator corresponds to a PDW, the $\eta$ mode can be seen as a PDW mode. It is depicted on Fig.\ \ref{fig:order-parameter}. Another way of picturing this $\eta$ mode is to say that it changes simultaneously the amplitudes of both the charge and the superconducting orders. As such, it can be seen as two amplitudons related by the constraint in Eq.\ \eqref{eq:su2-condition}.

We start by transforming Eq.\ \eqref{eq:lagrangian}:
\begin{align}
\mathcal{L} = &\sum_{a<b} \frac{\chi}{2} (n_a \partial_t n_b - n_b \partial_t n_a)^2 \nonumber
\\
&- \sum_{a<b} \frac{\rho_s}{2} (n_a \nabla n_b - n_b \nabla n_a)^2 - V(\textbf{n})
\label{eq:lagrangian-expanded}
\end{align}
by expanding the first term:
\begin{align}
\sum_{a<b} \frac{\chi}{2} (n_a \partial_t n_b - n_b \partial_t n_a)^2
= \sum_{a<b} \frac{\chi}{2} \left[ n_a^2 (\partial_t n_b)^2 \right. \nonumber
\\
\left. + n_b^2 (\partial_t n_a)^2 - 2 n_a \partial_t n_a n_b \partial_t n_b \right]
\end{align}
Since the norm of $\textbf{n}$ is fixed, we can use the two identities:
\begin{equation}
\textbf{n}^2 = 1 \text{ and } \textbf{n} \cdot \partial_t \textbf{n} = 0
\end{equation}
which enable us to simplify the first term to:
\begin{equation}
\sum_{a<b} \frac{\chi}{2} (n_a \partial_t n_b - n_b \partial_t n_a)^2
= \sum_{b} \frac{\chi}{2} (\partial_t n_b)^2
\end{equation}
Using the same procedure for the second term of Eq.\ \eqref{eq:lagrangian-expanded} gives us the following $O(3)$ non-linear $\sigma$ model:
\begin{align}
\mathcal{L} = &\sum_{b} \left[ \frac{\chi}{2} ( \partial_t n_b )^2 - \frac{\rho_s}{2} ( \nabla n_b )^2 \right] - V(\textbf{n})
\end{align}
In order to study the collective modes in this model, we place ourselves in the pure superconducting state $n_1$:
\begin{equation}
\mathcal{L} = \sum_{b \neq 1} \left[ \frac{\chi}{2} ( \partial_t n_b )^2 - \frac{\rho_s}{2} ( \nabla n_b )^2 \right] - V(\textbf{n})
\end{equation}
Minimising with respect to the two other coordinates $n_2$ and $n_3$, and setting $\chi = 1$, gives us equations corresponding to two collective modes:
\begin{align}
\partial_t^2 n_2 &= \rho_s \nabla^2 n_2 + (m_{CO} - m_{SC}) n_2 \label{eq:SC-mode1}
\\
\partial_t^2 n_3 &= \rho_s \nabla^2 n_3
\label{eq:SC-mode2}
\end{align}
The first one corresponds to rotating from $n_1$ to $n_2$, i.e. from superconductivity (which is the ground state we are in) to charge order: it is the $\eta$ mode. If $m_{SC} < m_{CO}$, it becomes massive. If $m_{SC} > m_{CO}$, the mass is negative, which is normal since we expanded around $n_1$, which in this case is not the ground state. The second collective mode corresponds to rotating from $n_1$ to $n_3$, i.e.\ from one superconducting coordinate to the other: it is the Goldstone boson of superconductivity. Note that it is massless.

We can also place ourselves in the pure charge ordered state $n_2$ and obtain a very similar non-linear $\sigma$ model:
\begin{equation}
\mathcal{L} = \sum_{b={1,3}} \left[ \frac{\chi}{2} ( \partial_t n_b )^2 - \frac{\rho_s}{2} ( \nabla n_b )^2 \right] - V(\textbf{n})
\end{equation}
Minimising with respect to the two other coordinates, and setting $\chi = 1$ gives us equations corresponding to two collective modes:
\begin{align}
\partial_t^2 n_1 &= \rho_s \nabla^2 n_1 + (m_{SC} - m_{CO}) n_1 \label{eq:CO-mode1}
\\
\partial_t^2 n_3 &= \rho_s \nabla^2 n_3 + (m_{SC} - m_{CO}) n_3 \label{eq:CO-mode2}
\end{align}
They correspond to rotating from $n_2$, the charge order coordinate, to $n_1$ and $n_3$ respectively, the two superconducting coordinate: both correspond to the $\eta$ mode. They are degenerate since the two superconducting coordinates are at the same energy. If $m_{SC} > m_{CO}$, both modes become massive. If $m_{SC} < m_{CO}$, charge order is no longer the ground state and therefore the two modes acquire negative masses.

One can calculate the renormalised parameter of a general version of the non-linear $\sigma$-model at finite temperature and applied magnetic field:
\begin{equation}
\mathcal{L} = \sum_{b=1}^3 \left[( \partial_t n_b )^2 - \rho_s( \nabla n_b )^2 - \kappa^2_0 n_b^2 \right]
\end{equation}
where $\kappa^2_0 = m_{SC} - m_{CO}$ is the bare anisotropy and $\rho_s$ is the stiffness. Note that here we consider a simply quadratic potential term $V$ which means that there is no uniform coexistence phase, and that the transitions towards superconductivity and charge order meet at a critical point on the zero-temperature axis. The renormalisation was done in a previous work \cite{Chakraborty2018}, and the renormalised anisotropy was calculated to be:
\begin{equation}
\kappa^2 = \kappa^2_0 \left( 1 - \frac{T}{2\pi \rho_s}\ln \left(\frac{\Lambda}{\sqrt{\kappa_0+B}} \right) \right)
\label{eq:kappa}
\end{equation}
where $\kappa_0$ is the bare anisotropy, $T$ the temperature, $\rho_s$ the stiffness associated with spatial variations of the order parameter, $\Lambda$ the upper momentum cutoff and $B$ the applied magnetic field. We show the evolution of $\kappa$ in the $B$-$T$ phase diagram in Fig.\ \ref{fig:mass}. The main feature of this evolution is that $\kappa$ softens when applying magnetic field, up to the transition to charge order, and then hardens again. It is zero at the quantum critical point at field $B=B_{CO}$, because the symmetry is rigorously valid there.

This anisotropy $\kappa$ is key since it enters all the equations for the $\eta$ modes as its mass. The softening of $\kappa$ pictured in Fig.\ \ref{fig:mass} under applied magnetic field therefore means that the mass of the collective modes will also soften when increasing magnetic field, up to the point where there is a transition to the charge ordered state, at which this mass will start hardening.

One can also, as stated above, consider a case where the phase diagram displays a coexistence phase. This coexistence corresponds to a PDW phase, and since the $\eta$ mode is a PDW mode, the coexistence phase is a condensation of the $\eta$ mode. Within this phase, the ground state is a mixture of superconducting and charge orders, and therefore the order parameter is not aligned with either axes. Its rotation towards pure superconducting and pure charge order states is massive, but this mass goes to zero at the tetracritical point where the symmetry is rigorously valid.

We can notice that the various equations that we derived for the $\eta$ mode in this section all have the same magnitude for the mass, namely the anisotropy $\kappa$. Moreover, the sign is also identical as long as we are deriving the modes close to the ground state of the system. Finally, the two modes derived close to $n_2$ (Eq.\ \eqref{eq:CO-mode1} and Eq.\ \eqref{eq:CO-mode2}) are identical. The propagator for the $\eta$ mode therefore is:
\begin{equation}
D(\textbf{q}, \omega) = \frac{1}{\omega^2 - \rho_s q^2 - \kappa^2} \label{eq:D propagator}
\end{equation}
where $\rho_s$ is the stiffness, $q$ the momentum close to $\textbf{Q}_0$ and $\omega$ the frequency. In the following, we use an anisotropy of 10 meV, a residual scattering of 2 meV and a temperature of 1 meV. A value for the stiffness was calculated in the framework of the eight hot-spots model studied in \cite{Meier13}. There, the stiffness obtained is $\rho_s = \frac{3\alpha}{16 \pi} \frac{T^*}{T}$ where $\alpha \approx 1$. If one sets the temperature $T$ to a tenth of the pseudogap temperature $T^*$, one obtains $\rho_s = 0.597$. We use this value in the following.

We have laid out the equations Eqs.\ \eqref{eq:SC-mode1} \eqref{eq:SC-mode2} \eqref{eq:CO-mode1} \eqref{eq:CO-mode2} for the collective modes present in the system. There are two types of collective modes: the rotation between the two coordinates of superconductivity and the $\eta$ mode corresponding to the fluctuation of the order parameter between charge order and superconductivity. We showed that the evolution of the mass of the $\eta$ mode, shown in Fig.\ \ref{fig:mass} in the case where there is no phase where charge order and superconductivity coexist, features a softening induced by magnetic field at the transition from superconductivity to the charge order. This is due to the fact that the SU(2) symmetry is verified exactly at this point of the phase diagram. We have shown that the mode corresponding to the rotation between charge and superconductivity corresponds to the $\eta$ operator defined in Eq.\ \eqref{eq:eta-operator}. We now turn to a self-consistent linear response formalism to study this mode in the weak-coupling limit.

\section{Weak coupling regime}
\label{Weak coupling regime}

In the previous section, we derived the shape of the $\eta$ mode in the strong coupling limit. Here we derive it in a weak coupling formalism. We follow closely the self-consistent linear response formalism previously used in the context of the negative-$U$ Hubbard model and of the SO(5) theory \cite{Demler1996, Demler1997}. The $\eta$ operator (Eq.\ \eqref{eq:eta-operator}) has charge 2, therefore we expect that the $\eta$ mode will only be excited within the superconducting phase. We will therefore consider a simple model taking only short-range antiferromagnetic correlations into account, and suppose that it has a superconducting ground state. We then apply a small perturbation on the density.

We start with the $t$-$J$ model \cite{MontielNeutrons2017} perturbed by an infinitesimal external field $\phi_\textbf{q}$ which couples to the density at momentum $-\textbf{q}$:
\begin{align}
H = &\sum_{\textbf{k} \sigma} \xi_\textbf{k} c^\dagger_{\textbf{k} \sigma} c_{\textbf{k} \sigma} \nonumber + \frac{1}{2}\rho_{-\textbf{q}} \phi_\textbf{q}
\\
&+ \frac{1}{2} \sum_{\textbf{k} \textbf{k}' \textbf{q}'} J(q') \sum_{\alpha \beta \gamma \delta} c^\dagger_{\textbf{k} \alpha} \vec{\sigma}_{\alpha \beta} c_{\textbf{k}+\textbf{q}' \beta} \vec{\sigma}_{\gamma \delta} c^\dagger_{\textbf{k}'+\textbf{q}' \gamma} c_{\textbf{k}' \delta}
\end{align}
where $\textbf{k}$, $\textbf{k}'$ and $\textbf{q}'$ are reciprocal space vectors, $\alpha$, $\beta$, $\gamma$ and $\delta$ are spin indices, $\xi_\textbf{k}$ is the electronic dispersion, $J$ is the antiferromagnetic coupling, $\vec{\sigma}$ is the vector of Pauli matrices and $\rho_\textbf{q}$ is the density:
\begin{align}
\rho_{\textbf{q} \uparrow} = \sum_{\textbf{k}} c^\dagger_{\textbf{k} + \textbf{q} \uparrow} c_{\textbf{k} \uparrow}
\\
\rho_{\textbf{q} \downarrow} = \sum_{\textbf{k}} c^\dagger_{\textbf{k} + \textbf{q} \downarrow} c_{\textbf{k} \downarrow}
\end{align}
We want to study the response at $\textbf{q}$ close to the charge ordering wave-vector $\textbf{Q}_0$, which is the vector linking two hot-spots. We use the completeness relation of the Pauli matrices:
\begin{equation}
\vec{\sigma}_{\alpha \beta} \vec{\sigma}_{\gamma \delta} = 2 \delta_{\alpha \delta} \delta_{\beta \gamma} - \delta_{\alpha \beta} \delta_{\gamma \delta}
\end{equation}
which gives us the following coupling term:
\begin{align}
\frac{1}{2} \sum_{\textbf{k} \textbf{k}' \textbf{q}'} J(q') \sum_{\alpha \beta} \left( c^\dagger_{\textbf{k} \alpha} c_{\textbf{k}+\textbf{q}' \beta} c^\dagger_{\textbf{k}'+\textbf{q}' \beta} c_{\textbf{k}' \alpha} \right. \nonumber
\\
\left. + c^\dagger_{\textbf{k} \alpha} c_{\textbf{k}+\textbf{q}' \alpha} c^\dagger_{\textbf{k}'+\textbf{q}' \beta} c_{\textbf{k}' \beta} \right)
\end{align}
We want to study the influence of the external field $\phi_\textbf{q}$ on the system in the superconducting state. In order to do this, we assume that this field excites a set of fields at the excitation momentum $\textbf{q}$. We go beyond standard random phase approximation methods by both considering the influence of particle-particle and particle-hole excitations on the response to $\phi_\textbf{q}$.

We assume a BCS superconducting ground state, as well as the fact that the expectation values for density and pairing at the excitation wave-vector $\textbf{q}$ are non-vanishing:
\begin{align}
u_\textbf{k}v_\textbf{k} &= \left\langle c_{-\textbf{k}\downarrow} c_{\textbf{k}\uparrow} \right\rangle = \left\langle c^\dagger_{\textbf{k}\uparrow} c^\dagger_{-\textbf{k}\downarrow} \right\rangle
\\
\hat{\eta}^*_\textbf{q} (t) &= \left\langle \eta^\dagger_\textbf{q} (t) \right\rangle
\\
\hat{\eta}_\textbf{q} (t) &= \left\langle \eta_\textbf{q} (t) \right\rangle
\\
\hat{\rho}_{\textbf{q} \uparrow} &= \left\langle \rho_{\textbf{q} \uparrow} \right\rangle
\\
\hat{\rho}_{\textbf{q} \downarrow} &= \left\langle \rho_{\textbf{q} \downarrow} \right\rangle
\end{align}
where the coefficients are given by:
\begin{align*}
u_\textbf{k}^2 &= \frac{1}{2} \left( 1 + \frac{\xi_\textbf{k}}{E_\textbf{k}} \right)
&
v_\textbf{k}^2 &= \frac{1}{2} \left( 1 - \frac{\xi_\textbf{k}}{E_\textbf{k}} \right)
\end{align*}
where $E_\textbf{k} = \sqrt{\xi_\textbf{k} + \Delta}$, $\xi_\textbf{k}$ is the electronic dispersion in the normal state and $\Delta$ is the superconducting gap which we set to be constant in this section for simplicity. We will consider a $d$-wave superconducting gap in the following section.

Linearising the Hamiltonian with respect to these expectation values gives:
\begin{align}
H = &\sum_{\textbf{k} \sigma} \xi_\textbf{k} c^\dagger_{\textbf{k} \sigma} c_{\textbf{k} \sigma} + \Delta \sum_\textbf{k} \left( c^\dagger_{\textbf{k} \uparrow} c^\dagger_{-\textbf{k} \downarrow} + c_{-\textbf{k} \downarrow} c_{\textbf{k} \uparrow} \right) \nonumber
\\
& - 6 J_0 \hat{\eta}^*_\textbf{q} \sum_\textbf{k} c_{-\textbf{k} \downarrow} c_{\textbf{k}+\textbf{q} \uparrow} - 6 J_0 \hat{\eta}_\textbf{q} \sum_\textbf{k} c^\dagger_{\textbf{k}-\textbf{q} \uparrow} c^\dagger_{-\textbf{k} \downarrow} \nonumber
\\
& - 6 J_0 \left( \hat{\rho}_{\textbf{q} \uparrow} + \frac{\phi_\textbf{q}}{2} \right) \sum_\textbf{k} c_{\textbf{k}-\textbf{q} \downarrow} c_{\textbf{k} \downarrow} \nonumber
\\
&- 6 J_0 \left( \hat{\rho}_{\textbf{q} \downarrow} + \frac{\phi_\textbf{q}}{2} \right) \sum_\textbf{k} c_{\textbf{k}-\textbf{q} \uparrow} c_{\textbf{k} \uparrow} 
\label{eq:linearised-Hamiltonian}
\end{align}
where $J_0$ is the antiferromagnetic coupling close to a vector linking two hot-spots, and where, without loss of generality, we rescaled $\phi_\textbf{q}$ by a factor $6 J_0$ to make Eq.\ \ref{eq:linearised-Hamiltonian} simpler. Here we consider not only the particle-hole channel as usually done, but also the particle-particle channel corresponding to the $\eta$ and $\eta^\dagger$ operators, which is the part that we are most interested in. We obtain the same expression as the one obtained in the case of the negative-U Hubbard model \cite{Demler1996}, with a $6 J_0$ coefficient instead of $U$. We obtain a coefficient 6 here due to the fact that we did not include the $n_i n_j$ term in the Hamiltonian \cite{Lee06}. We assume the BCS self-consistency relation:
\begin{equation}
\Delta = 6 J_0 \sum_\textbf{k} u_\textbf{k} v_\textbf{k}
\end{equation}

We consider the first line of $H$, which is a BCS Hamiltonian, as the ground state Hamiltonian $H_0$. The rest is proportional to fields which are proportional to $\phi_\textbf{q}$, and we will treat it as a perturbation $H_1$. The response is given by the Kubo formula:
\begin{equation}
\langle O(t) \rangle = -i \int_{-\infty}^t dt' \left\langle \left[ O(t), H_1 (t') \right] \right\rangle_{H_0}
\end{equation}
where the operator $O$ can be any of the operators $\eta^\dagger_\textbf{q}$, $\eta_\textbf{q}$, $\rho_{\textbf{q} \uparrow}$ or $\rho_{\textbf{q} \downarrow}$. We define the frequency-dependent operators by: $O(\omega) = \int dt e^{i \omega t} O(t)$. Applying this to the operators in our problem gives:
\begin{align}
\hat{\eta}^*_\textbf{q} (\omega) = & 6 J_0 \sum_{\textbf{k} \nu} G_\textbf{k}(\nu) G_{\textbf{k} + \textbf{q}} (-\nu-\omega) \hat{\eta}^*_\textbf{q} (\omega) \nonumber
\\
&- 6 J_0 \sum_{\textbf{k} \nu} F^\dagger_\textbf{k}(\nu) F^\dagger_{\textbf{k} + \textbf{q}} (\nu+\omega) \hat{\eta}_\textbf{q} (\omega) \nonumber
\\
&- 6 J_0 \sum_{\textbf{k} \nu} F^\dagger_\textbf{k}(\nu) G_{\textbf{k} + \textbf{q}} (\nu-\omega) \left( \hat{\rho}_\textbf{q} (\omega) + \phi_\textbf{q} (\omega) \right)
\label{eq:equation-eta-dagger}
\end{align}
\begin{align}
\hat{\eta}_\textbf{q} (\omega) = & - 6 J_0 \sum_{\textbf{k} \nu} F_\textbf{k}(\nu) F_{\textbf{k} + \textbf{q}} (\nu+\omega) \hat{\eta}^*_\textbf{q} (\omega) \nonumber
\\
&+ 6 J_0 \sum_{\textbf{k} \nu} G_\textbf{k}(\nu) G_{\textbf{k} + \textbf{q}} (\omega-\nu) \hat{\eta}_\textbf{q} (\omega) \nonumber
\\
&- 6 J_0 \sum_{\textbf{k} \nu} F_\textbf{k}(\nu) G_{\textbf{k} + \textbf{q}} (\nu+\omega) \left( \hat{\rho}_\textbf{q} (\omega) + \phi_\textbf{q} (\omega) \right)
\label{eq:equation-eta}
\end{align}
\begin{align}
\hat{\rho}_\textbf{q} (\omega) = & - 6 J_0 \sum_{\textbf{k} \nu} F_{\textbf{k} + \textbf{q}}(\nu) G_\textbf{k} (\nu-\omega) \hat{\eta}^*_\textbf{q} (\omega) \nonumber
\\
&- 6 J_0 \sum_{\textbf{k} \nu} F^\dagger_{\textbf{k} + \textbf{q}}(\nu) G_\textbf{k} (\nu+\omega) \hat{\eta}_\textbf{q} (\omega) \nonumber
\\
&+ 6 J_0 \sum_{\textbf{k} \nu} \left[ F_\textbf{k}(\nu) F_{\textbf{k} + \textbf{q}} (\nu+\omega) \right. \nonumber
\\
&+ \left. G_\textbf{k}(\nu) G_{\textbf{k}+\textbf{q}} (\nu+\omega) \right] \left( \hat{\rho}_\textbf{q} (\omega) + \phi_\textbf{q} (\omega) \right)
\label{eq:equation-rho}
\end{align}
where $G$ is the normal propagator, $F$ and $F^\dagger$ are the anomalous propagators, and $\nu$ and $\omega$ are fermionic and bosonic Matsubara frequencies respectively. The calculation of these coefficients follows exactly what was done previously in \cite{Demler1996}. For completeness, we reproduce the corresponding diagrams in the Appendices: Fig.\ \ref{fig:appendix-diagrams-eta-dagger}, \ref{fig:appendix-diagrams-eta} and \ref{fig:appendix-diagrams-rho}. The results can be summarised in the following matrix equation:
\begin{equation}
\begin{pmatrix}
\hat{\eta}^*_{\textbf{q}\omega} + \hat{\eta}_{\textbf{q}\omega} \\
\hat{\eta}^*_{\textbf{q}\omega} - \hat{\eta}_{\textbf{q}\omega} \\
\rho_{\textbf{q}\omega}
\end{pmatrix}
=
\begin{pmatrix}
t_{++} & t_{+-} & m_+ \\
t_{+-} & t_{--} & m_- \\
m_+ & m_- & 6 J_0 \chi_{bcs}
\end{pmatrix}
\begin{pmatrix}
\hat{\eta}^*_{\textbf{q}\omega} + \hat{\eta}_{\textbf{q}\omega} \\
\hat{\eta}^*_{\textbf{q}\omega} - \hat{\eta}_{\textbf{q}\omega} \\
\rho_{\textbf{q}\omega} + \phi_{\textbf{q}\omega}
\end{pmatrix}
\label{eq:matrix system}
\end{equation}
where the matrix coefficients are:
\begin{widetext}
\begin{align}
t_{++} = & -6 J_0 \sum_\textbf{k} \left[ 1 - n_F(E_{\textbf{k}+\textbf{q}}) - n_F(E_{\textbf{k}}) \right] \frac{A_{\textbf{k}\textbf{q}} (u_\textbf{k} u_{\textbf{k}+\textbf{q}} - v_\textbf{k} v_{\textbf{k}+\textbf{q}})^2}{\omega^2 - A^2_{\textbf{k}\textbf{q}}}
\nonumber \\
& - 6 J_0 \sum_\textbf{k} \left[ n_F(E_{\textbf{k}+\textbf{q}}) - n_F(E_{\textbf{k}}) \right] \frac{B_{\textbf{k}\textbf{q}} (u_\textbf{k}v_{\textbf{k}+\textbf{q}} + v_\textbf{k} u_{\textbf{k}+\textbf{q}} )^2}{\omega^2 - B^2_{\textbf{k}\textbf{q}}}
\label{eq:t++}
\end{align}
\begin{align}
t_{+-} = & 6 J_0 \sum_\textbf{k} \left[ 1 - n_F(E_{\textbf{k}+\textbf{q}}) - n_F(E_{\textbf{k}}) \right] \frac{\omega (1-v^2_{\textbf{k}+\textbf{q}}-v^2_\textbf{k})}{\omega^2 - A^2_{\textbf{k}\textbf{q}}}
\nonumber \\
& + 6 J_0 \sum_\textbf{k} \left[ n_F(E_{\textbf{k}+\textbf{q}}) - n_F(E_{\textbf{k}}) \right] \frac{\omega ( v^2_{\textbf{k}+\textbf{q}} -v^2_\textbf{k})}{\omega^2 - B^2_{\textbf{k}\textbf{q}}}
\label{eq:t+-}
\end{align}
\begin{align}
t_{--} = & -6 J_0 \sum_\textbf{k} \left[ 1 - n_F(E_{\textbf{k}+\textbf{q}}) - n_F(E_{\textbf{k}}) \right] \frac{A_{\textbf{k}\textbf{q}} (u_\textbf{k}u_{\textbf{k}+\textbf{q}} + v_\textbf{k} v_{\textbf{k}+\textbf{q}} )^2}{\omega^2 - A^2_{\textbf{k}\textbf{q}}}
\nonumber \\
& - 6 J_0 \sum_\textbf{k} \left[ n_F(E_{\textbf{k}+\textbf{q}}) - n_F(E_{\textbf{k}}) \right] \frac{B_{\textbf{k}\textbf{q}} (u_\textbf{k}v_{\textbf{k}+\textbf{q}} - v_\textbf{k} u_{\textbf{k}+\textbf{q}} )^2}{\omega^2 - B^2_{\textbf{k}\textbf{q}}}
\label{eq:t--}
\end{align}
\begin{align}
m_+ = & -6 J_0 \sum_\textbf{k} \left[ 1 - n_F(E_{\textbf{k}+\textbf{q}}) - n_F(E_{\textbf{k}}) \right] \frac{c_{\textbf{k}\textbf{q}} (u_\textbf{k} v_{\textbf{k}} + u_{\textbf{k}+\textbf{q}} v_{\textbf{k}+\textbf{q}} )}{\omega^2 - A^2_{\textbf{k}\textbf{q}}}
\nonumber \\
& + 6 J_0 \sum_\textbf{k} \left[ n_F(E_{\textbf{k}+\textbf{q}}) - n_F(E_{\textbf{k}}) \right] \frac{c_{\textbf{k}\textbf{q}} (u_{\textbf{k}+\textbf{q}} v_{\textbf{k}+\textbf{q}} - v_\textbf{k} u_\textbf{k} )}{\omega^2 - B^2_{\textbf{k}\textbf{q}}}
\label{eq:m+}
\end{align}
\begin{align}
m_- = & 6 J_0 \sum_\textbf{k} \left[ 1 - n_F(E_{\textbf{k}+\textbf{q}}) - n_F(E_{\textbf{k}}) \right] \frac{\omega (u_\textbf{k} v_{\textbf{k}} + u_{\textbf{k}+\textbf{q}} v_{\textbf{k}+\textbf{q}} )}{\omega^2 - A^2_{\textbf{k}\textbf{q}}}
\nonumber \\
& - 6 J_0 \sum_\textbf{k} \left[ n_F(E_{\textbf{k}+\textbf{q}}) - n_F(E_{\textbf{k}}) \right] \frac{\omega (u_{\textbf{k}+\textbf{q}} v_{\textbf{k}+\textbf{q}} - u_\textbf{k} v_\textbf{k} )}{\omega^2 - B^2_{\textbf{k}\textbf{q}}}
\label{eq:m-}
\end{align}
\begin{align}
\chi_{bcs} = & - \sum_\textbf{k} \left[ 1 - n_F(E_{\textbf{k}+\textbf{q}}) - n_F(E_{\textbf{k}}) \right] \frac{A_{\textbf{k}\textbf{q}} (u_{\textbf{k}+\textbf{q}} v_{\textbf{k}} + u_\textbf{k} v_{\textbf{k}+\textbf{q}} )^2}{\omega^2 - A^2_{\textbf{k}\textbf{q}}}
\nonumber \\
& + \sum_\textbf{k} \left[ n_F(E_{\textbf{k}+\textbf{q}}) - n_F(E_{\textbf{k}}) \right] \frac{B_{\textbf{k}\textbf{q}} (u_\textbf{k} u_{\textbf{k}+\textbf{q}} - v_\textbf{k} v_{\textbf{k}+\textbf{q}})^2}{\omega^2 - B^2_{\textbf{k}\textbf{q}}}
\end{align}
\end{widetext}
where $n_F$ is the Fermi-Dirac distribution, $A_{\textbf{k}\textbf{q}} = E_{\textbf{k}+\textbf{q}} + E_\textbf{k}$, $B_{\textbf{k}\textbf{q}} = E_{\textbf{k}+\textbf{q}} - E_\textbf{k}$, and $c_{\textbf{k}\textbf{q}} = \xi_{\textbf{k}+\textbf{q}} + \xi_\textbf{k}$.

\begin{figure}
\centering
\includegraphics[width=5cm]{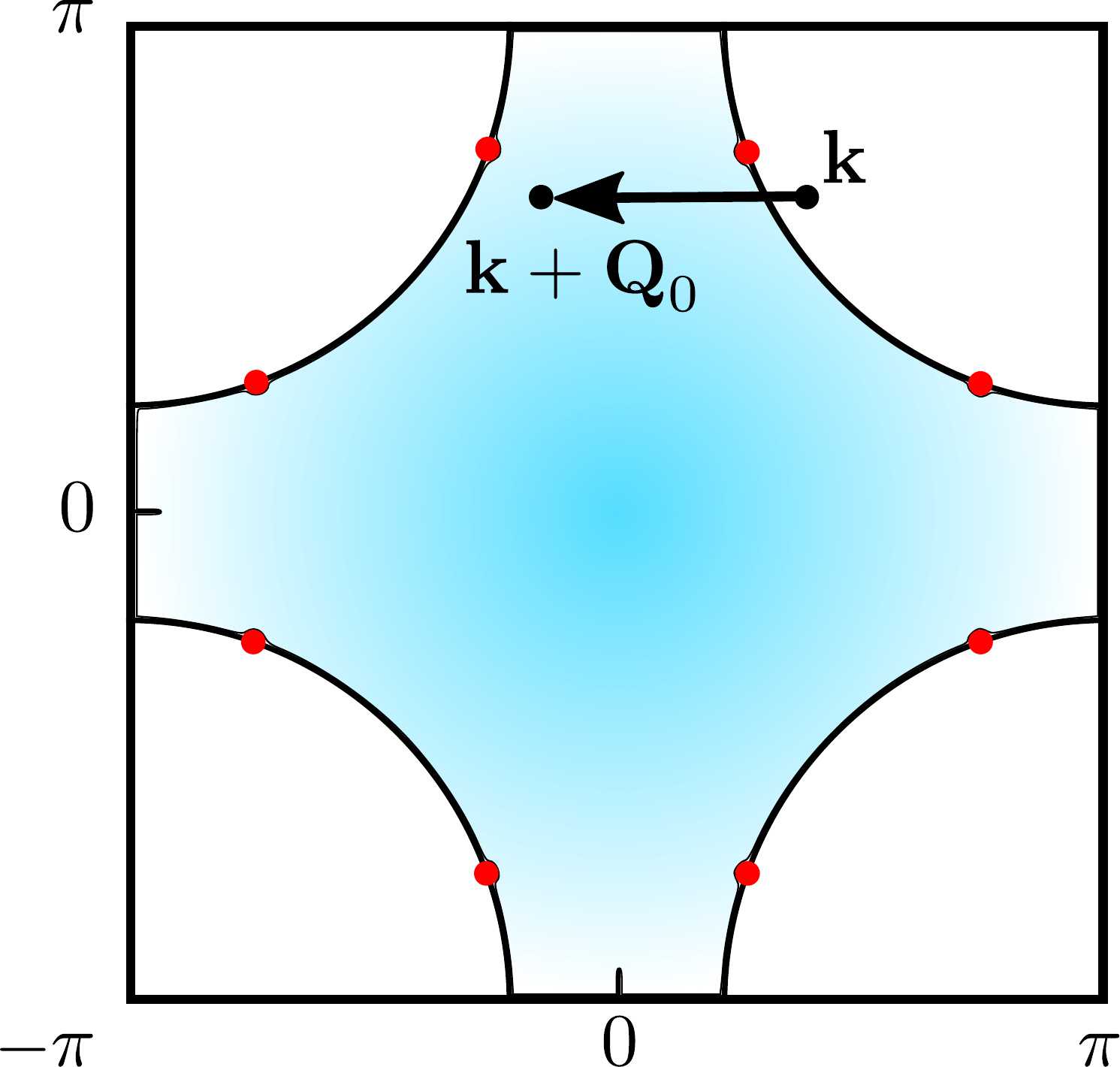}
\caption{We approximate the electronic dispersion by its variation close to the hot-spots: $\xi_\textbf{k} = v_F k_\parallel + \frac{k^2_\perp}{m^*}$. When going from $\textbf{k}$ to $\textbf{k}+\textbf{Q}_0$, one reverses the component of the dispersion parallel to the Fermi surface but the component perpendicular to the Fermi surface does not change. This gives $c_{\textbf{k} \textbf{Q}_0} \approx 2 \frac{k^2_\perp}{m^*}$.}
\label{fig:approximation} 
\end{figure}

We now have a system of coupled equations (Eq.\ \eqref{eq:matrix system}) which we want to solve in order to find expressions for $\hat{\eta}^*_{\textbf{q}\omega}$ and $\hat{\eta}_{\textbf{q}\omega}$. In order to do this we make two approximations: first, we set $T=0$ and thus $n_F(E_{\textbf{k}+\textbf{q}}) = n_F(E_{\textbf{k}}) = 0$. This causes in particular all the coefficients of the matrix in Eq.\ \eqref{eq:matrix system} to reduce to their first term. We notice that in Eq.\ \eqref{eq:t+-} and Eq.\ \eqref{eq:m-}, $\omega$ can be taken out of the sum on $\textbf{k}$. Moreover we know from the case of the Hubbard model \cite{Demler1996} that:
\begin{equation}
A_{\textbf{k}\textbf{q}}  (u_\textbf{k} u_{\textbf{k}+\textbf{q}} - v_\textbf{k} v_{\textbf{k}+\textbf{q}})^2 = c_{\textbf{k}\textbf{q}} (1-v^2_{\textbf{k}+\textbf{q}}-v^2_\textbf{k})
\end{equation}
which changes the coefficient of Eq.\ \eqref{eq:t++}. In order to simplify these coupled equations we want to take $c_{\textbf{k}\textbf{q}}$ out of the sums in Eq.\ \eqref{eq:t++} and Eq.\ \eqref{eq:m+}, similarly to what we did for $\omega$.

We focus on the momentum $\textbf{q}=\textbf{Q}_0$, and approximate the electronic dispersion for $\textbf{k}$ close to the hot-spots: $\xi_\textbf{k} = v_F k_\parallel + \frac{k^2_\perp}{m^*}$ where $k_\parallel$ and $k_\perp$ are the components of $\textbf{k}$ parallel and perpendicular to the Fermi surface respectively, the $v_F$ is the Fermi velocity and $m^*$ is the curvature of the dispersion close to the hot-spots. This gives us: $c_{\textbf{k} \textbf{Q}_0} \approx 2 \frac{k^2_\perp}{m^*}$, as can be seen from plotting these two points on the Fermi surface on Fig.\ \ref{fig:approximation}. Knowing that the maximal value of $\frac{k^2_\perp}{m^*}$ is much smaller than the one of $v_F k_\parallel$, we approximate $2\frac{k^2_\perp}{m^*}$ by its maximum which we name $\alpha$. Because we approximate $c_{\textbf{k} \textbf{Q}_0}$ by a constant, we can now take it out of the sums in Eq.\ \eqref{eq:t++} and Eq.\ \eqref{eq:m+}.

We start by summing the first row in Eq.\ \eqref{eq:matrix system} times $\omega$ and the second row times $\alpha$. We use the identity:
\begin{equation}
A_{\textbf{k}\textbf{q}} (1-v^2_{\textbf{k}+\textbf{q}}-v^2_\textbf{k}) = c_{\textbf{k}\textbf{q}} (u_\textbf{k} u_{\textbf{k}+\textbf{q}} + v_\textbf{k} v_{\textbf{k}+\textbf{q}})^2
\end{equation}
to simplify the term $\alpha \times t_{--}$ and obtain:
\begin{equation}
\omega \left( \hat{\eta}^*_{\textbf{Q}_0\omega} + \hat{\eta}_{\textbf{Q}_0\omega} \right) = \left[ - \alpha + 6 J_0 (1 - n) \right] \left( \hat{\eta}^*_{\textbf{Q}_0\omega} - \hat{\eta}_{\textbf{Q}_0\omega} \right)
\label{eq:intermediate 1}
\end{equation}
where $n=2\sum_\textbf{k} v_\textbf{k}^2$ is the density. We want to transform the third equation in Eq.\ \eqref{eq:matrix system}. We start by transforming our expression for $\chi_{bcs}$ by using the identity:
\begin{align}
A_{\textbf{k}\textbf{q}} (v_\textbf{k} u_{\textbf{k}+\textbf{q}} + u_\textbf{k} v_{\textbf{k}+\textbf{q}})^2 = &\left( \frac{A^2_{\textbf{k}\textbf{q}}-\omega^2}{2 \Delta} + \frac{\omega^2 - c^2_{\textbf{k}\textbf{q}}}{2 \Delta} \right) \nonumber
\\
&\times (u_\textbf{k} v_\textbf{k} + u_{\textbf{k}+\textbf{q}} v_{\textbf{k}+\textbf{q}})
\end{align}
and obtain:
\begin{equation}
\chi_{bcs} = - \sum_\textbf{k} \frac{u_\textbf{k} v_{\textbf{k}} + u_{\textbf{k}+\textbf{q}} v_{\textbf{k}+\textbf{q}}}{\omega^2 - A^2_{\textbf{k}\textbf{q}}} \frac{\omega^2 - c^2_{\textbf{k}\textbf{q}}}{2 \Delta } + \frac{1}{6 J_0}
\label{eq:chi-bcs-transformed}
\end{equation}
We add to the third row of Eq.\ \eqref{eq:matrix system} times $2\Delta$ with this expression for $\chi_{bcs}$ the first row times $\alpha$ and the second row times $\omega$, which gives:
\begin{align}
\omega \left( \hat{\eta}^*_{\textbf{Q}_0\omega} - \hat{\eta}_{\textbf{Q}_0\omega} \right) = &\left[ - \alpha + 6 J_0 (1 - n) \right] \left( \hat{\eta}^*_{\textbf{Q}_0\omega} + \hat{\eta}_{\textbf{Q}_0\omega} \right)
\nonumber \\
&+ 2\Delta \phi_{\textbf{Q}_0 \omega}
\label{eq:intermediate 2}
\end{align}
The two equations \eqref{eq:intermediate 1} and \eqref{eq:intermediate 2} form a system. Their sum and difference give us respectively:
\begin{align}
\hat{\eta}^*_{\textbf{Q}_0 \omega} &= \frac{\Delta}{\omega + \kappa} \phi_{\textbf{Q}_0 \omega}
\label{eq:eta-dagger-weak-coupling}
\\
\hat{\eta}_{\textbf{Q}_0 \omega} &= -\frac{\Delta}{\omega - \kappa} \phi_{\textbf{Q}_0 \omega}
\label{eq:eta-weak-coupling}
\end{align}
where
\begin{equation}
\kappa = \alpha - 6 J_0(1-n)
\end{equation}
is the resonance energy. These two equations describe the $\eta$ mode which corresponds to the rotation between the charge and superconducting order parameters, as discussed in the previous section. Similarly to the previous section, in Eqs.\ \eqref{eq:eta-dagger-weak-coupling} and \eqref{eq:eta-weak-coupling} the resonance diverges at a finite resonance energy $\kappa$, which is the weak-coupling equivalent of the $\kappa$ defined in the strong coupling section in Eq.\ \eqref{eq:kappa}. In the strong coupling model we took the magnetic field into account, and obtained the magnetic field dependence of $\kappa$. Here we did not take the magnetic field in account, but directly linked $\kappa$ to the electronic density $n$. It is 1 at half-filling, and then decreases when the system is hole-doped. Therefore $\kappa$ decreases when the hole-doping increases.

Eq.\ \ref{eq:eta-dagger-weak-coupling} and Eq.\ \ref{eq:eta-weak-coupling} are proportional to $\Delta$, meaning that they are only finite in the superconducting state. This is in line with the fact that because these operators bear charge 2 they can only be excited in the superconducting condensate, where there is an electron-hole symmetry.

In this section we have derived the shape of the $\eta$-mode in the weak coupling limit using a self-consistent linear response formalism. We found that it diverges at the charge ordering wave-vector $\textbf{Q}_0$ and at the resonance energy $\kappa$, similarly to what was obtained in the previous section in the strong coupling formalism. It can be contrasted to what was obtained in the SO(5) theory, based on a composite order parameter rotating from antiferromagnetism to charge order \cite{Demler2004}. There, the mode diverges at the antiferromagnetic ordering wave-vector $\left( \pi,\pi \right)$ and at a finite energy which goes to zero at around 5\% doping, where the system goes from antiferromagnetism to superconductivity.

In the following sections, we discuss how the $\eta$ mode can be seen in the imaginary part of the charge susceptibility.

\section{Strong coupling between charge and superconducting order parameters: charge susceptibility}
\label{Strong coupling: charge susceptibility}

In Sec.\ \ref{Strong coupling regime} and Sec.\ \ref{Weak coupling regime}, we have derived the $\eta$ mode at momentum $\textbf{Q}_0$ using strong and weak coupling formalisms. Here we use the strong coupling formalism to derive the influence of the $\eta$ mode on the imaginary part of the charge susceptibility. In particular, we want to find the resonant contribution of this mode, meaning the contribution peaked at the resonance frequency and at the resonance wave-vector. 

The central feature of our strong coupling model is the definition of a composite order parameter which embraces both charge and superconducting orders. In Sec.\ \ref{Strong coupling regime}, we discussed how this naturally causes the arising of the $\eta$ mode corresponding to the rotation between these two orders. This bosonic mode is unique to this model, in that it only exists because charge order and superconductivity are gathered in a single order parameter $\textbf{n}$ and linked by the SU(2) symmetry. As such, it forms the most direct signature of the SU(2) symmetry and therefore of this model.

Because it corresponds to the rotation between charge order and superconductivity, the $\eta$ mode is charge 2. This can be seen by the fact that, for example, the operator linking $n_1$ and $n_2$ is $L_{12} = i\frac{\eta^\dagger - \eta}{2}$. The operators $\eta^\dagger$ and $\eta$ indeed bear two electronic charges. This mode therefore acts on particle pairs, which means that it will be much more visible in the superconducting state.

The bare polarisation bubble in a superconductor can be put in this form \cite{Schrieffer64, Norman07}:
\begin{align}
\chi_{bcs} (\omega, \textbf{q}) = \sum_\textbf{k} \left[ \left( u_\textbf{k} u_{\textbf{k}+\textbf{q}} + v_\textbf{k} v_{\textbf{k}+\textbf{q}} \right)^2 \frac{n_F(E_{\textbf{k}+\textbf{q}}) - n_F(E_\textbf{k})}{\omega - E_{\textbf{k}+\textbf{q}} + E_\textbf{k}}  \right. \nonumber
\\
+ \left( u_\textbf{k} v_{\textbf{k}+\textbf{q}} - v_\textbf{k} u_{\textbf{k}+\textbf{q}} \right)^2 \frac{1 - n_F(E_{\textbf{k}+\textbf{q}}) - n_F(E_\textbf{k})}{\omega + E_{\textbf{k}+\textbf{q}} + E_\textbf{k}} \nonumber
\\
+ \left. \left( u_\textbf{k} v_{\textbf{k}+\textbf{q}} - v_\textbf{k} u_{\textbf{k}+\textbf{q}} \right)^2 \frac{n_F(E_{\textbf{k}+\textbf{q}}) + n_F(E_\textbf{k}) - 1}{\omega - E_{\textbf{k}+\textbf{q}} - E_\textbf{k}} \right]
\label{eq:BCS-chi0}
\end{align}
It is the bare contribution of the electronic sector to the charge susceptibility in a superconductor. In the presence of an interaction it can be renormalised, as done for example in the random phase approximation formalism \cite{Norman07}.

Here we want to study the contribution of the $\eta$ mode to the charge susceptibility. It will sum up with other contributions, such as the one from the electrons of Eq.\ \eqref{eq:BCS-chi0}. We place ourselves in the superconducting state in the Nambu formalism which gives us the usual two normal and two anomalous propagators. 

\begin{figure}
\centering
\includegraphics[width=8cm]{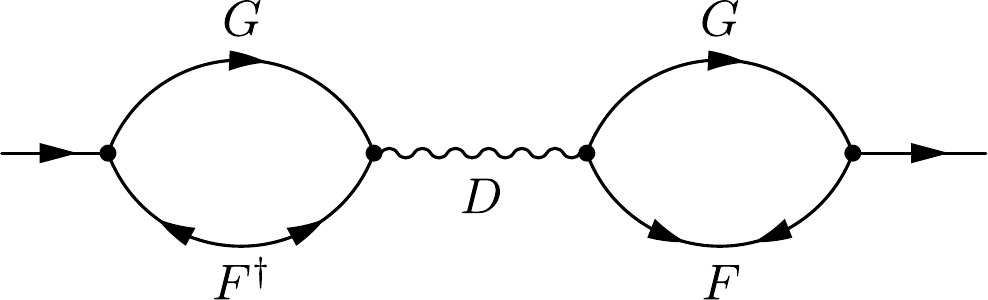}
\caption{Diagram for the resonant contribution of the $\eta$ mode to the charge susceptibility in the superconducting phase $\chi_\eta$. $D$ is the bosonic propagator corresponding to the $\eta$ mode. Note that the two bottom propagators are anomalous propagators, non-vanishing only in the superconducting phase.}
\label{fig:diagram} 
\end{figure}

We name the main resonant contribution to the charge susceptibility $\chi_\eta$ in the superconducting phase, and represent it by the diagram in Fig.\ \ref{fig:diagram}. This contribution of the collective mode to the charge susceptibility only appears in the superconducting phase, and we therefore expect an sharp change at the superconducting transition. The full diagram $\chi_\eta$ (Fig.\ \ref{fig:diagram}) is given by:
\begin{equation}
\chi_\eta(\textbf{q}, \omega) = |B(\textbf{q}, \omega)|^2 D(\textbf{q}, \omega) \label{eq:chiB}
\end{equation}
where $B$ is the contribution of the fermions, and $D$ is the bosonic propagator.

\subsection{Fermionic polarisation}
\label{Fermionic polarisation}

\begin{figure*}
\centering
\includegraphics[width=18cm]{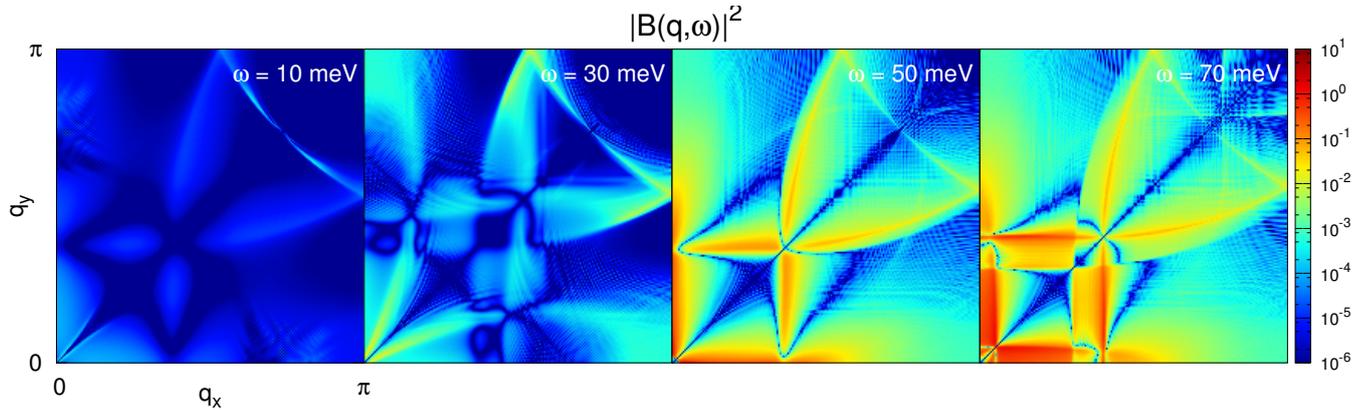}
\caption{$|B|^2$ in the Brillouin zone at four different frequencies. The red colour is for $B$ positive and the blue colour is for $B$ negative. Note that $B$ has $d$-wave symmetry, since an anomalous propagator enters its expression (Eq.\ \eqref{eq:B-short}). $|B|^2$ is much larger above 50 meV, due to the fact that the Stoner continuum is approximately at this energy for the momenta with the largest response.}
\label{fig:continuum} 
\end{figure*}

\begin{figure}
\centering
\includegraphics[width=6cm]{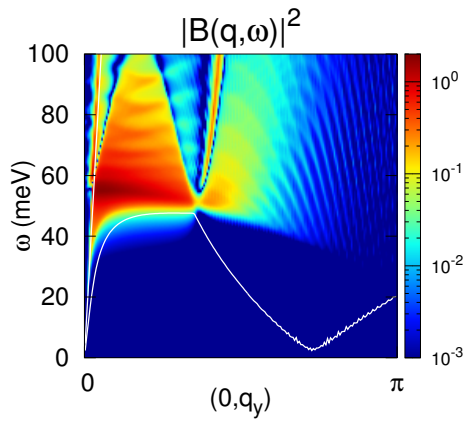}
\caption{$|B|^2$ as a function of frequency along the axis of the Brillouin zone. The white solid lines are the top and bottom of the Stoner continuum. $|B|^2$ is maximum on a straight line at about 50 eV.}
\label{fig:continuum-energy} 
\end{figure}

Here we calculate the left-hand side polarisation of $\chi_\eta$ (Fig.\ \ref{fig:diagram}), which we name $B(\textbf{q}, \omega)$. We discuss its evolution in frequency on the axis of the Brillouin zone, its evolution in momentum, and finally how it changes with the location in the phase diagram. We start by giving its expression:
\begin{align}
B(\textbf{q}, \omega) = & \sum_{\textbf{k} \epsilon} F^\dagger(\textbf{k}, \epsilon) \nonumber
\\
&\times \left[ G(-\textbf{k}+\textbf{q}, -\epsilon+\omega) - G(-\textbf{k}+\textbf{q}, \epsilon-\omega) \right]
\label{eq:B-short}
\end{align}
where $G$ and $F^\dagger$ are normal and anomalous propagators respectively, $\textbf{k}$ and $\textbf{q}$ are reciprocal space vectors, $\epsilon$ and $\omega$ are fermionic and bosonic frequencies respectively. The two terms in Eq.\ \eqref{eq:B-short} correspond to the two Nambu indices. Since the second term is the opposite of the complex conjugate of the first, their sum is twice their imaginary part. This is a consequence of the symmetry between the propagators in the two Nambu indices: $G_{11}(\omega) = -G^*_{22}(-\omega)$. Because $F^\dagger$ is an anomalous propagator, $B=0$ outside of the superconducting phase. Summing over fermionic frequencies $\epsilon$ gives:
\begin{widetext}
\begin{align}
B(\textbf{q}, \omega) = & 2\sum_{\textbf{k}} u_{-\textbf{k}+\textbf{q}}^2 u_\textbf{k} v_\textbf{k} \text{Im} \left( \frac{ 1- n_F(E_\textbf{k}) - n_F(E_{-\textbf{k}+\textbf{q}}) }{i\omega - E_{-\textbf{k}+\textbf{q}} - E_\textbf{k}} + \frac{n_F(E_{-\textbf{k}+\textbf{q}}) - n_F(E_\textbf{k})}{i\omega - E_{-\textbf{k}+\textbf{q}} + E_\textbf{k}}
\right) \nonumber
\\
&+ 2\sum_{\textbf{k}} v_{-\textbf{k}+\textbf{q}}^2 u_\textbf{k} v_\textbf{k} \text{Im} \left(\frac{n_F(E_{-\textbf{k}+\textbf{q}}) - n_F(E_\textbf{k})}{i\omega + E_{-\textbf{k}+\textbf{q}} - E_\textbf{k}} + \frac{1-n_F(E_\textbf{k}) - n_F(E_{-\textbf{k}+\textbf{q}}) }{i\omega + E_{-\textbf{k}+\textbf{q}} + E_\textbf{k}} \right) \label{eq:B-long}
\end{align}
\end{widetext}
where $n_F$ is the Fermi-Dirac distribution, $E_\textbf{k} = \sqrt{\xi_\textbf{k} + \Delta_\textbf{k}}$, $\xi_\textbf{k}$ is the electronic dispersion in the normal state, $\Delta_\textbf{k}$ is the superconducting $d$-wave gap and $\text{Im}$ is the imaginary part. We use a dispersion $\xi_\textbf{k}$ obtained from angle-resolved photoemission data in \cite{Norman07}: $\xi_\textbf{k} = t_0 + t_1/2 \left[ \cos(k_x) + \cos(k_y) \right] + t_2 \cos(k_x) \cos(k_y) + t_3 /2 \left[ \cos(2 k_x) + \cos(2 k_y)) \right] + t_4 /2 \left[ \cos(2 k_x) \cos(k_y) + \cos(k_x) \cos(2 k_y) \right] + t_5 \cos(2 k_x) \cos(2 k_y)$ where the hopping parameters are:  $t_0 = 0.1305$ eV, $t_1 = -0.5951$ eV, $t_2 = 0.1636$ eV, $t_3 = -0.0519$ eV, $t_4 = -0.1117$ eV and $t_5 = 0.0510$ eV. It corresponds to a hole doping of 17\% \cite{Norman:1995dd}, and to a distance between hot-spots of $Q_x = Q_y = 1.12$ in reciprocal lattice units, which means that it is not close to a van Hove singularity.

One can notice that $B$ is formed of products of one normal propagator with one anomalous propagator. This is unlike $\chi_{bcs}$ (Eq.\ \eqref{eq:BCS-chi0}) which is formed of products of either two normal or two anomalous propagators. From the numerators of the four terms in Eq.\ \eqref{eq:B-long} we can already see that the dominant contribution of $B$ will come from the first term. Indeed, at zero temperature, the Fermi functions vanish and we are left only with the first and the last term. Moreover, for an infinitesimal scattering, these terms are Dirac functions at $\omega = E_{-\textbf{k}+\textbf{q}} + E_\textbf{k}$ and $\omega = -E_{-\textbf{k}+\textbf{q}} - E_\textbf{k}$ respectively. Given that $E_\textbf{k} > 0$, the first term dominates for positive frequencies and will only be sizable above the lower bound of the Stoner continuum: $\min_\textbf{k} \left( E_{-\textbf{k}+\textbf{q}} + E_\textbf{k} \right)$. We discuss the physical interpretation of the shape of this line in details in Appendix \ref{Decomposition of the Stoner continuum}.

We show the frequency evolution of $|B(\textbf{q}, \omega)|^2$ along the axis of the Brillouin zone on Fig.\ \ref{fig:continuum-energy}. It is clearly delimited by the lower bound of the Stoner continuum: $\min_\textbf{k} \left( E_{-\textbf{k}+\textbf{q}} + E_\textbf{k} \right)$, which is plotted on the same figure. Most of the weight is condensed on a very flat line around 50 meV and close to the origin of the Brillouin zone. This energy is of the order of twice the superconducting gap. Indeed, given that $\textbf{Q}_0$ links two hot-spots, if $\textbf{k}$ is at a hot-spot, we get $\xi_{-\textbf{k}+\textbf{Q}_0} = \xi_\textbf{k} = 0$, and $\min_\textbf{k} \left( E_{-\textbf{k}+\textbf{Q}_0} + E_\textbf{k} \right) = 2 \Delta_{\textbf{Q}_0}$.

The evolution of $|B(\textbf{q}, \omega)|^2$ in the Brillouin zone is displayed on Fig.\ \ref{fig:continuum}. It is $d$-wave, which follows logically from the fact that the superconducting parameter $\Delta_\textbf{k}$ is $d$-wave, and that it is a prefactor of the anomalous propagator $F^\dagger$ and therefore also of $B$ (Eq.\ \eqref{eq:B-short}).

We now discuss the evolution of $|B(\textbf{q}, \omega)|^2$ in the phase diagram. Since the maximum weight of $|B(\textbf{q}, \omega)|^2$ is located close to the lower boundary of the Stoner continuum, we can focus our discussion on this boundary. Outside of the superconducting phase, the lower boundary of the Stoner continuum is zero. Inside the superconducting phase, it gets larger when the gap grows. For example, the boundary of the Stoner continuum at $\textbf{Q}_0$ is strictly proportional to the gap: $\min_\textbf{k} \left( E_{-\textbf{k}+\textbf{Q}_0} + E_\textbf{k} \right) = 2 \Delta_{\textbf{Q}_0}$. We illustrate the variation of $|B(\textbf{q}, \omega)|^2$ with the size of the superconducting gap in details in Fig.\ \ref{fig:bubble-delta} in Appendix \ref{Evolution of the continuum bubble with the size of the superconducting gap}.

\subsection{Contribution of the $\eta$ mode to the charge susceptibility}

\begin{figure*}
\centering
\includegraphics[width=18cm]{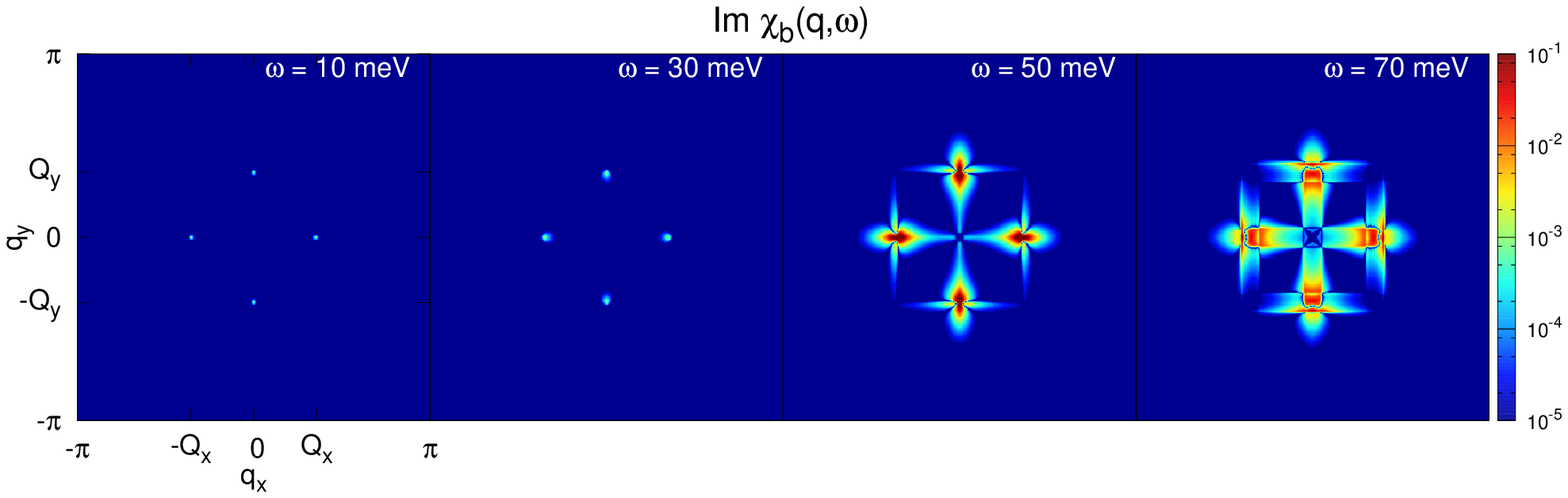}
\caption{Imaginary part of the contribution of the collective mode to the charge susceptibility $\chi_\eta$, plotted in the Brillouin zone for four different frequencies. It peaks at the charge ordering wave vectors $\textbf{Q}_0$, situated on the axes. Note that we are considering the same logarithmic scale for all four plots, as indicated on the right hand side. This shows that for a set frequency, the peak in $\chi_\eta$ is always situated close to $Q_0$, and that the strength of this peaks grows larger when crossing the Stoner continuum, at about 50 meV. Moreover, the texture of the continuum $B$ is clearly visible, and much smaller in magnitude.}
\label{fig:bubble} 
\end{figure*}

\begin{figure*}
\centering
\includegraphics[width=18cm]{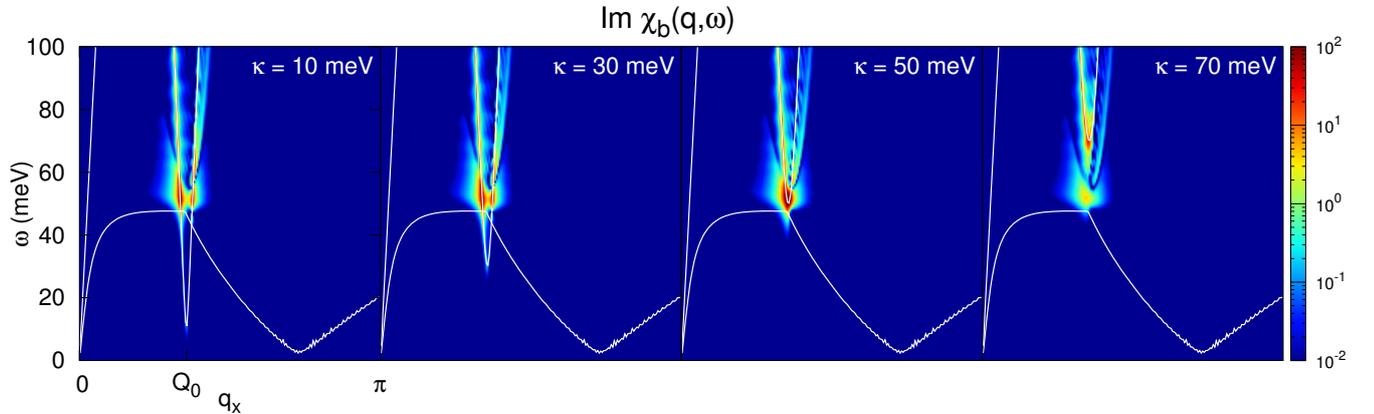}
\caption{Frequency dependence of the imaginary part of $\chi_\eta$ along the axis of the Brillouin zone, for a range of bosonic masses. We use the same logarithmic scale for all four plots, as indicated on the right hand side. In grey are also plotted the bottom and top of the Stoner continuum, along with the line where the bosonic propagator diverges. Note that for low masses, the response peaks at two spots just above the continuum bottom line. These two spots merge when the mass is raised. The structure of the continuum $B$, as well as the fennel of the bosonic mode, are visible and lower in magnitude.}
\label{fig:mass-evolution} 
\end{figure*}

Here we calculate the imaginary part of the total contribution of the $\eta$ mode to the charge susceptibility $\text{Im}( \chi_\eta )$. We then discuss its evolution in the Brillouin zone, in frequency, and its dependence on the mass of the collective mode. Finally, we examine the magnitude of the response.

The evolution of $\text{Im}( \chi_\eta )$ in the Brillouin zone at several frequencies is represented in Fig.\ \ref{fig:bubble}. We can see that the response is precisely located at the resonance energy at the charge ordering wave vectors. Away from the resonance energy, the response disperses but with a much smaller magnitude, and follows the shape of $B$ (Fig.\ \ref{fig:continuum}), i.e.\ the shape of the electronic dispersion, which is also responsible for specific patterns in, for example, the spin response \cite{Norman07}. We discuss the link between the evolution of $B$ and $\text{Im}( \chi_\eta )$ in the Brillouin zone in Appendix \ref{Influence of the continuum bubble}.

The variation of $\text{Im}( \chi_\eta )$ in frequency along the axis of the Brillouin zone is displayed on Fig.\ \ref{fig:mass-evolution}. The response is limited to the region between the two continuum boundaries, and therefore is zero below the Stoner continuum bottom edge, as shown on Fig.\ \ref{fig:mass-evolution}. This is due to the fact that $|B(\textbf{q}, \omega)|^2$ is a factor of $\text{Im}( \chi_\eta )$, and therefore of the fact that $\text{Im}( \chi_\eta )$ can only be observed in the superconducting state, and inside the Stoner continuum. The mass of the mode $\kappa$ can therefore be smaller than the peaks appearing in $\text{Im}( \chi_\eta )$.

\begin{figure}
\centering
\includegraphics[width=8cm]{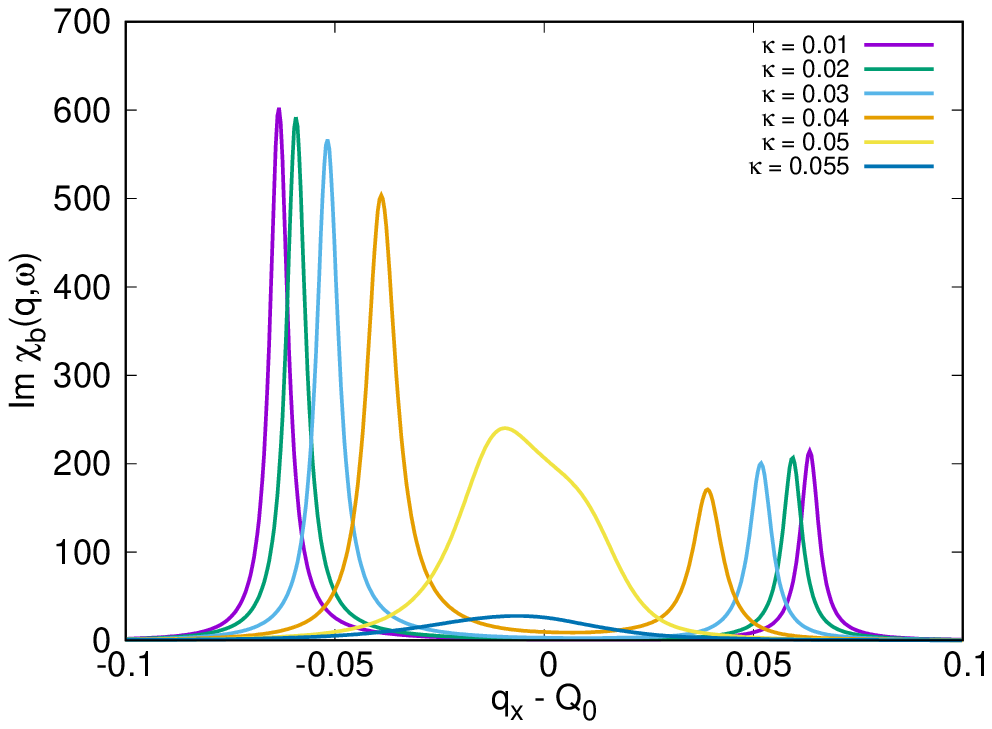}
\caption{Momentum dependence of the imaginary part of $\chi_\eta$ along the axis of the Brillouin zone, at $\omega$ = 50 meV, for a range of bosonic masses. This frequency corresponds to the edge of the  continuum, as plotted on Fig.\ \ref{fig:continuum-energy}. Note that for low masses, the response consists in two peaks. When the mass is raised, these two peaks get closer and finally merge at a mass close to the frequency of the edge of the continuum. When the mass is raised further, the peak becomes much smaller.}
\label{fig:mass-evolution-50meV} 
\end{figure}

The mass of the collective mode depends, as discussed above, on the temperature and the applied magnetic field, but it is also material and doping-dependent. We display the evolution of the response $\text{Im}( \chi_\eta )$ for different masses in Fig.\ \ref{fig:mass-evolution}. The collective mode disperses following a large slope, as discussed in Sec.\ \ref{Collective mode}. The two branches of the dispersion are therefore close together. For $\kappa$ lower than the continuum edge, the dispersion of the bosonic mode crosses the continuum edge twice. The response peaks close to these two points, and extends above, following the bosonic divergence line. The response is maximal for $\kappa$ close to the continuum edge, and consists of only one peak. For $\kappa$ higher than the continuum edge, the response is mostly following the parabolic bosonic divergence line, with some extra weight close to the continuum edge where $B$ is largest. The overall response is much weaker than for the two previous cases. In order to track the separation of these two peaks more precisely, on Fig.\ \ref{fig:mass-evolution-50meV} we plot the evolution of $\text{Im}( \chi_\eta )$ along the axis of the Brillouin zone at $\omega = 50$ meV for different masses. We see clearly that for $\kappa$ of the order of the continuum edge the two peaks merge into one, which collapses when the mass increases more.

Finally, Fig.\ \ref{fig:mass-evolution-50meV} allows us to clearly see the magnitude of the peaks at 50 meV. We notice that the peak closer to the center of the Brillouin zone is about twice as large as the one closer to the edge of the Brillouin zone. We can compare the contribution to the imaginary part of the charge susceptibility from the $\eta$ mode to the bare contribution from the superconducting condensate $\chi_{bcs}$. The maximum of the latter is of the order of 1, and is therefore two orders of magnitude smaller than the contribution of the $\eta$ mode $\chi_\eta$. Moreover, given that $\chi_\eta$ is only visible inside the superconducting phase, it is possible to isolate it experimentally by substracting the susceptibilities inside and outside the superconducting phase.

\subsection{Evolution in the phase diagram}

We now discuss the evolution of $\text{Im}( \chi_\eta )$ in the phase diagram of the cuprates: first its evolution with magnetic field, then with doping and temperature, and finally its relationship to other observables.

In Sec.\ \ref{Strong coupling regime}, we showed that $\kappa$, the mass of the $\eta$ mode, softens at the transition between charge and superconducting orders under magnetic field. This softening is due to the fact that at the transition magnetic field, the SU(2) symmetry between these two orders is exact. The evolution of $\chi_\eta$ with magnetic field allows us to track the softening of $\kappa$, which is a unique feature of our model: the signature of the SU(2) symmetry. At low field, $\kappa$ is large and $\text{Im}( \chi_\eta )$ features either one or two nearby peaks, depending on the exact value of $\kappa$. When increasing the applied magnetic field, $\kappa$ is lowered and the peak in $\text{Im}( \chi_\eta )$ splits in two peaks which separate more and more when increasing the applied magnetic field. The middle of the two peaks remains centred on the charge order wave vector $\textbf{Q}_0$.

The resonance is centred on $\textbf{Q}_0$, which connects two hot-spots. It therefore follows the evolution of the electronic dispersion with doping: when increasing doping, one gets closer to the van Hove singularity, and $\textbf{Q}_0$ gets smaller. It has been measured that in some compounds the van Hove singularity coincides with the end of the pseudogap \cite{Doiron-Leyraud2017}. In our model, this is visible when measuring the resonance, which converges to zero momentum at high doping when the van Hove singularity coincides with the end of the pseudogap. Moreover, we know experimentally that 12\% doping is the doping at which charge order is the closest, i.e.\ where we need to apply less magnetic field to reach it. In our model, this means that it is the doping at which $\kappa$ is minimal at zero field. The distance between the two peaks in $\text{Im}( \chi_\eta )$ at zero field should therefore be minimal at that doping.

When increasing the temperature while staying inside the superconducting phase, the superconducting gap is lowered. As discussed in Section \ref{Fermionic polarisation}, when the superconducting gap is lowered, the lower boundary of the Stoner continuum goes to lower energies. Therefore the frequency at which the $\eta$ mode is observed in $\text{Im}( \chi_\eta )$ also gets smaller. This could mean that the charge susceptibility would feature a soft mode at a finite wave vector, indicating the arising of a charge density wave order. However the magnitude of $B$ will also go to zero at $T_c$, which will prevent the $\eta$ mode from causing such a condensation. Moreover, we stress that the evolution of $\kappa$ with temperature is very steep, as plotted in Fig.\ \ref{fig:mass}. Experimentally, the softening of $\kappa$ close to $T_c$ will therefore be hard to measure.

Finally, the charge susceptibility, beyond its direct connection to the MEELS response, also enters other important observables. In particular, it renormalises the electron-phonon coupling \cite{Bruus2004book}. Therefore the resonance we calculated in the charge susceptibility can cause the electron-phonon coupling to diverge at specific momenta and to enhance the phonon response measured by probes such as RIXS. This is in line with RIXS measurements, which found that the resonance displayed on Fig.\ \ref{fig:RIXS} was peaked on a known phonon line in this material \cite{Chaix2017,Zhou2018}. If this RIXS resonance is indeed due to the renormalisation of the electron-phonon coupling by the $\eta$ mode, its evolution in temperature will not be the same as the one of the susceptibility: it should remain at the same frequency above $T_c$, and decrease in intensity. Lastly, since the $\eta$ mode softens, its crossing with the phonon line should happen at momenta further away from $\textbf{Q}_0$, so the resonance momentum should move slightly away from $\textbf{Q}_0$.

In this section we derived the influence of the $\eta$ mode, corresponding to the rotation from superconductivity to charge order, on the imaginary part of the charge susceptibility inside the superconducting phase in the strong coupling formalism. We found that it peaks just above the continuum edge. For high $\kappa$ it peaks only at one point, while for $\kappa$ below the continuum edge it splits in two peaks which separate further and further when $\kappa$ decreases. Since we showed that $\kappa$ decreases with applied magnetic field (Fig.\ \ref{fig:mass}), the peaks split when applying magnetic field (Fig.\ \ref{fig:mass-evolution-50meV}), while their middle always remaining centred on the charge ordering wave vector $\textbf{Q}_0$. This splitting is maximum at the transition between charge and superconducting orders, which is a signature of the SU(2) symmetry which is exact there. We now check the consistency between strong and weak coupling formalisms and discuss how this resonance is seen in the charge susceptibility in the weak coupling formalism, developed in Sec.\ \ref{Weak coupling regime}.

\section{Weak coupling: Charge susceptibility}

Here we derive the charge susceptibility in a weak coupling formalism. In Sec.\ \ref{Weak coupling regime}, we have derived a system of equations Eq.\ \eqref{eq:matrix system} which enabled us to calculate expressions for $\hat{\eta}^*_{\textbf{Q}_0 \omega}$ and $\hat{\eta}_{\textbf{Q}_0 \omega}$ in Eq.\ \eqref{eq:eta-dagger-weak-coupling} and Eq.\ \eqref{eq:eta-weak-coupling}. We now input these expressions into the third row of Eq.\ \eqref{eq:matrix system}. This directly gives us the charge susceptibility:
\begin{equation}
\chi(\textbf{Q}_0,\omega) = \frac{\rho_{\textbf{Q}_0 \omega}}{\phi_{\textbf{Q}_0 \omega}} = \chi_1(\textbf{Q}_0,\omega) + \chi_2(\textbf{Q}_0,\omega)
\label{eq:chi-weak}
\end{equation}
where
\begin{align}
\chi_1(\textbf{Q}_0,\omega) &= - \frac{2 \chi_{bcs}}{1-6J_0 \chi_{bcs}}
\label{eq:chi-weak-1} 
\\
\chi_2(\textbf{Q}_0,\omega) &= - \frac{4\Delta J_\omega}{1-6 J_0 \chi_{bcs}} \frac{\omega^2 + \alpha \kappa}{\omega^2 - \kappa^2}
\label{eq:chi-weak-2}
\end{align}
and
\begin{equation}
J_\omega = \sum_\textbf{k} \frac{u_\textbf{k} v_\textbf{k} + u_{\textbf{k}+\textbf{Q}_0} v_{\textbf{k}+\textbf{Q}_0}}{\omega^2 - A^2_{\textbf{k} \textbf{Q}_0}}
\end{equation}
The obtained charge susceptibility is strickingly the sum of a renormalised BCS susceptibility: $\chi_1(\textbf{Q}_0,\omega)$ and of a specific contribution due to the presence of the finite expectation values for the $\eta$ operators: $\chi_2(\textbf{Q}_0,\omega)$. This second contribution has a factor $J_\omega$ which has poles of at $A_{\textbf{k} \textbf{Q}_0} = E_{\textbf{k}+\textbf{Q}_0} + E_\textbf{k}$, which is exactly the definition of the Stoner continuum as discussed in the previous section.

We take Eq.\ \eqref {eq:chi-bcs-transformed} at $\textbf{q} = \textbf{Q}_0$ and use the approximation, discussed in Sec.\ \ref{Weak coupling regime}, that $c_{\textbf{k}\textbf{Q}_0} = \alpha$ is a constant to obtain:
\begin{equation}
1 - 6 J_0 \chi_{bcs} = 6 J_0 \frac{\omega^2-\alpha^2}{2 \Delta} J_\omega
\label{eq:chi-bcs-identity}
\end{equation}
which we input in the denominator of Eq.\ \eqref{eq:chi-weak-2}, giving:
\begin{equation}
\chi_2(\textbf{Q}_0,\omega) = - \frac{4\Delta^2}{3 J_0 (\omega^2-\alpha^2)} \frac{\omega^2 + \alpha \kappa}{\omega^2 - \kappa^2}
\end{equation}
We decompose this fraction into fractions with single poles and isolate the part which diverges at $\kappa$:
\begin{equation}
\chi(\textbf{Q}_0,\omega) = \frac{\Delta^2}{9 J_0^2 (1-n)} \frac{1}{\omega-\kappa} + \chi_{rest}(\textbf{Q}_0,\omega)
\label{eq:weak-coupling-chi}
\end{equation}
where $\chi_{rest}(\textbf{Q}_0,\omega)$ is the sum of the contributions which do not diverge at $\kappa$ which in particular includes $\chi_1$.

The result in Eq.\ \eqref{eq:weak-coupling-chi} does not feature poles at $A_{\textbf{k} \textbf{Q}_0}$, which define the Stoner continuum, unlike Eq.\ \eqref{eq:chi-weak-2} where $J_\omega$ and $\chi_{bcs}$ enter the equation. These two contributions canceled following the use of the identity Eq.\ \eqref{eq:chi-bcs-identity}. The self-consistency imposed by the method kills this Stoner continuum. We believe that the self-consistence imposes too strict a condition at this stage.

The strength of the response is proportional to $\Delta^2$, as shown in Eq.\ \eqref{eq:weak-coupling-chi} and Eq.\ \eqref{eq:B-long}. Therefore in this formalism, similarly to what we calculated in the previous section, the resonance is only present in the superconducting state.

Here we calculated the charge susceptibility in the weak coupling formalism developed in Sec.\ \ref{Weak coupling regime}. Similarly to what we calculated in the strong coupling regime in the previous section, the charge susceptibility diverges at the charge ordering wave-vector $\textbf{Q}_0$ and resonance energy $\kappa$ due to the contribution of the $\eta$ mode. This contribution scales with the square of the superconducting gap, which means that it is only present in the superconducting phase.

\section{Discussion of experiments}
\label{Discussion of experiments}

In the previous sections, we discussed the resonance arising from the excitation of the $\eta$ mode, corresponding to the rotation between charge and superconducting orders, and its resonant contribution to the imaginary part of the charge susceptibility. Here we discuss how to experimentally access this resonance. Since it is charge two, spin zero and at finite momentum, we turn to charge probes which can access finite $\textbf{q}$: momentum-resolved electron energy-loss spectroscopy (MEELS) and resonant X-ray scattering (RIXS). We also discuss the case of Raman scattering and optical conductivity.

\subsection{Momentum-resolved energy-electron loss spectroscopy}

MEELS was developed recently by combining electron energy-loss spectroscopy (EELS) with angular aligment techniques used in neutron or X-ray scattering experiments \cite{Vig2017}. It directly probes the imaginary part of the charge susceptibility. The use of EELS enables to probe the charge response with a very good energy resolution: it was claimed that for a beam energy of 50 eV the resolution in energy is below 0.5 meV \cite{Vig2017}. Using these alignment techniques led to achieving a resolution in momentum of 0.06 \AA$^{-1}$ \cite{Vig2017}. Since the beam energy has to be small for achieving good energy resolution, MEELS uses a reflection geometry, which makes it a surface-sensitive probe, similarly to ARPES. Finally, it can probe both normal and superconducting states, and is therefore particularly interesting for the study of superconductors.

Two MEELS experiments have already been performed, unfortunately neither can address the topic of the collective mode yet. The first measurement was done as a proof of concept in the paper presenting the details of the experimental technique \cite{Vig2017}. Spectra at finite energies between zero and 120 meV were measured, but they are dominated by the contribution coming from the diagonal, which strongly obscures the features for example on the axes of the Brillouin zone. This measurement was however performed on Bi$_2$Sr$_2$CaCu$_2$O$_{8+x}$ (Bi2212) whose samples are known to feature dislocations along the $(\pi,\pi)$ direction \cite{Borisenko2004}, which could be responsible for the large response along the diagonal. More detailed work was subsequently performed on the same material but focused on the detection of a plasmonic resonance close to 1 eV \cite{Mitrano2017}.

\begin{figure}
\centering
\includegraphics[width=8cm]{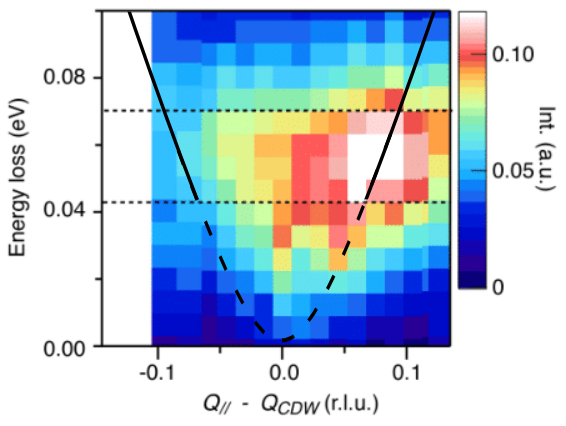}
\caption{Resonance close to $Q_0$ measured by RIXS on an underdoped Bi$_2$Sr$_2$CaCu$_2$O$_{8+\delta}$ (Bi2212) sample, of doping $p = 8-9 \%$ \cite{Chaix2017}. The black line is the dispersion of the collective mode fitted to this experiment. The black line is solid when inside the continuum, and dashed when outside of the continuum. Note that this fitting does not strongly constrains the value of the mass of the mode, but that it gives an order of magnitude for the stiffness of the mode.}
\label{fig:RIXS} 
\end{figure}

\subsection{Resonant X-ray scattering}
\label{Resonant X-ray scattering}

Comparisons between theory and experiment can already be drawn with recent resonant X-ray scattering results. RIXS measures the total electronic density susceptibility. This is different from the charge susceptibility since it includes all electrons, while the charge is only the difference between the densities of positive and negative charges. Indeed, a neutral material bears a zero charge but a finite electronic density. RIXS is therefore much more sensitive to displacements of large quantities of electrons, like plasmons, while MEELS is more sensitive to charge variations. Nevertheless, we expect RIXS to be sensitive to the collective mode of the model presented above since it bears a finite charge. So far, the energy resolution which has been achieved in RIXS experiment is around 40 meV, which is higher than the one in MEELS but still could enable one to see the influence of the collective mode on the electronic density response. Moreover, it can measure inside and outside of the superconducting phase, and therefore check our prediction that there is a change in the contribution of the $\eta$ mode at the superconducting transition.

A resonance at around 60 meV was detected using RIXS at doping $p = 8-9 \%$, along the axis of the Brillouin zone, slightly further away from the origin than the charge ordering wave vector \cite{Chaix2017,Zhou2018}. It is shown on Fig.\ \ref{fig:RIXS}. This peak has been interpreted as the signature of a phonon and modeled in the normal state \cite{Devereaux2016}. However it has not yet been seen using other experimental techniques and is weakened outside the superconducting state.

This resonance does correspond in energy and momentum to the response derived in the previous sections. Fitting this RIXS resonance to the bosonic dispersion for the $\eta$ mode calculated in Sec.\ \ref{Collective mode} does not give us much information about the mass of the collective mode, since there is only one point to be fitted. But it can give us information on the order of magnitude of its stiffness. For a mass of 10 meV, we obtain a stiffness of 0.597, while for zero mass we obtain a stiffness of 0.617. The latter result is plotted on Fig.\ \ref{fig:RIXS} along with the experimental data. These fitted value are very close to the value of the stiffness obtained theoretically in Sec.\ \ref{Collective mode}.

Time-resolved resonant soft x-ray scattering has also been performed on La$_{2-x}$Ba$_x$CuO$_4$ just above the superconducting transition temperature and found overdamped excitations at very low energies, below 2 meV \cite{Mitrano2018}. Such a low excitation energy corresponds to the mass of the collective mode derived in the previous sections. Moreover the very high damping could be due to the fact that the measurement was conducted outside of the superconducting phase.

\subsection{Raman scattering and optical conductivity}

Raman scattering detected a resonance at 41 meV in the $A_{1g}$ channel at optimal doping \cite{Cooper1988, Staufer1992, Sacuto1997, Gallais04, LeTacon05b}. It is not seen in either the $B_{1g}$ or $B_{2g}$ channels. This $A_{1g}$ resonance is measured at doping up to the edge of the pseudogap phase, and down to 12\%. This means that it is not found at the doping where the RIXS experiment detailed in the previous section was performed. It was shown that long-range Coulomb interactions screen the 2$\Delta$ contribution to the $A_{1g}$ channel, meaning that it cannot be simply explained by the presence of the superconducting condensate.

\begin{figure}
\centering
\includegraphics[width=8cm]{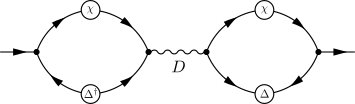}
\caption{The Raman response was derived in the context of the SU(2) theory, in the case where charge order and superconductivity coexist, using this diagram \cite{Montiel15a}. There are two anomalous propagators corresponding to the superconducting order, and two anomalous propagators corresponding to the charge order. The latter shift the resonance at $\textbf{Q}_0$ to zero momentum, and it was shown that it appears in the $A_{1g}$ channel of the Raman response, similarly to the experimental observations.}
\label{fig:diagram-Raman} 
\end{figure}

The Raman response is more difficult to derive, since it measures the charge susceptibility at $\textbf{q}=0$, and that we want to obtain the resonant contribution of the $\eta$ mode which is peaked at $\textbf{Q}_0$. We therefore need to shift this momentum to $\textbf{q}=0$. One solution that was explored was to insert a charge order in the model, but this had the inconvenient to impose the coexistence between charge and superconducting orders \cite{Montiel15a}. In that case, the presence of long-range charge order at $\textbf{Q}_0$ shifts the resonance to zero momentum, as shown on Fig.\ \ref{fig:diagram-Raman}, and the resonance appears in the $A_{1g}$ channel of the Raman response \cite{Montiel15a}. Other solutions which don't necessitate coexistence exist, in particular by considering disorder or the presence of phonons. They would apply a drag to the collective mode, meaning that the observed frequency would be the sum of the ones of the phonon and the collective mode, and that the response function would be brought back to $\textbf{q}=0$. These possibilities will be explored in a future publication.

Finally, the $\eta$ mode could be seen experimentally in probes that are sensitive to all types of fluctuations. This is in particular the case of optical conductivity: it was shown recently that the amplitude mode in a superconductor could be excited by applying a small ac electric field in the presence of a supercurrent \cite{Moor2017}. This could be used to detect other collective mode, such as the $\eta$ mode studied here, also in relationship with the phonon drag mechanism described in the previous paragraph. Optical conductivity measurements could in particular probe our prediction that the contribution of the $\eta$ mode must change at $T_c$, by substracting measurements on the two sides of the transition.

Here, we limited ourselves to studying the imaginary part of the charge susceptibility, in order to make the connection with the MEELS experiment discussed above. The calculation of the Raman and optical conductivity responses for this model will be done in a different work.

\section{Conclusion}

Here we derived the influence of the fluctuation of the SU(2) order parameter on the charge susceptibility in the superconducting phase. First we used a strong coupling formalism to build the composite order parameter which includes both superconducting and charge order components and has a fixed length beneath the pseudogap transition temperature $T^*$. Then we wrote a Lie algebra which rotates components onto one another. This enabled us to build a model corresponding to the rotation of this order parameter, and to obtain the shape of the collective mode corresponding to these fluctuations. The renormalisation of the mass of this mode at finite temperature evolves with applied magnetic field. In particular, we find that the mass of the collective mode softens at the transition between charge and superconducting orders. This is a unique feature of the symmetry between these two orders which becomes exact at this point of the phase diagram. It is even more remarkable given that this point can be reached by tuning two parameters which can be varied continuously experimentally, that is temperature and applied magnetic field. This is an advantage compared to the case, for example of the SO(5) symmetry, which becomes exact at a precise hole doping, which can be difficult to reach experimentally, if not impossible for specific families of cuprate materials.

We then studied the $\eta$ mode in the superconducting state in the weak coupling regime, by using a self-consistent linear response formalism. We showed that the $\eta$ mode diverge at the charge ordering wave-vector and at finite frequency.

We derived the resonant contribution of the $\eta$ mode to the charge susceptibility. Because this mode has charge 2, we considered a diagram in the superconducting state including anomalous propagators. We found that the response is peaked at the charge ordering wave-vector and at the frequency corresponding to the lower bound of the Stoner continuum, and that away from this frequency it disperses away from it. In order to study the charge susceptibility in the $B$-$T$ phase diagram, we studied the frequency-dependent response along the edge of the Brillouin zone for a range of masses. Above the lower bound of the Stoner continuum, when the mass is small, the response follows the two branches of the bosonic propagator and peaks twice just above the continuum line. These two peaks merge when the resonance frequency is raised and disappear when it grows much larger than the continuum line. The separation between these two peaks is magnetic field dependent and is maximal at high magnetic field, close to the transition towards the charge ordered phase. This is the signature of the SU(2) symmetry which is exact at this point of the phase diagram. This gives a strong prediction on the MEELS experimental results, that can be used as a test of the SU(2) theory. We calculated the charge susceptibility in the weak coupling regime, and found that it diverges at the charge ordering wave-vector, similarly to what was derived in the strong coupling regime.

In the last section, we discussed how the resonant contribution of the $\eta$ mode to the charge response can be seen experimentally, and focused on four types of experiments: MEELS, RIXS, Raman scattering and optical conductivity. In particular, MEELS can measure directly the imaginary part of the charge susceptibility, and therefore be directly compared to the results presented here. Moreover, we compared the stiffness obtained by theory to the one fitted to recent RIXS measurements, and found a very close agreement.

In this paper we derived the $\eta$ mode of the SU(2) theory and its influence on the charge susceptibility using two different formalism. We found that it is charge two, spin zero, corresponds to a PDW operator and that it peaks at the charge ordering wave vector. It disperses away from its resonance energy following a slope that corresponds very well to recent RIXS measurements. It can be seen in the superconducting phase in the imaginary part of the charge susceptibility, inside the continuum. Moreover, we derived the dependence of its mass with magnetic field and found a very specific evolution which is directly linked to the SU(2) symmetry. Our predictions can directly be compared to MEELS and RIXS measurements, and form an experimental test of the SU(2) theory.

\section*{Acknowledgements}

We would like to thank Peter Abbamonte, Laura Chaix, Hermann Freire, Yvan Sidis and Kejin Zhou for stimulating discussions. This work received financial support from the ERC, under grant agreement AdG-694651-CHAMPAGNE. The authors also  would like to thank the International Institute of Physics (Natal, Brazil), where parts of this work were done, for hospitality.

\appendix

\section{Influence of the continuum bubble}
\label{Influence of the continuum bubble}

\begin{figure*}
\centering
\includegraphics[width=18cm]{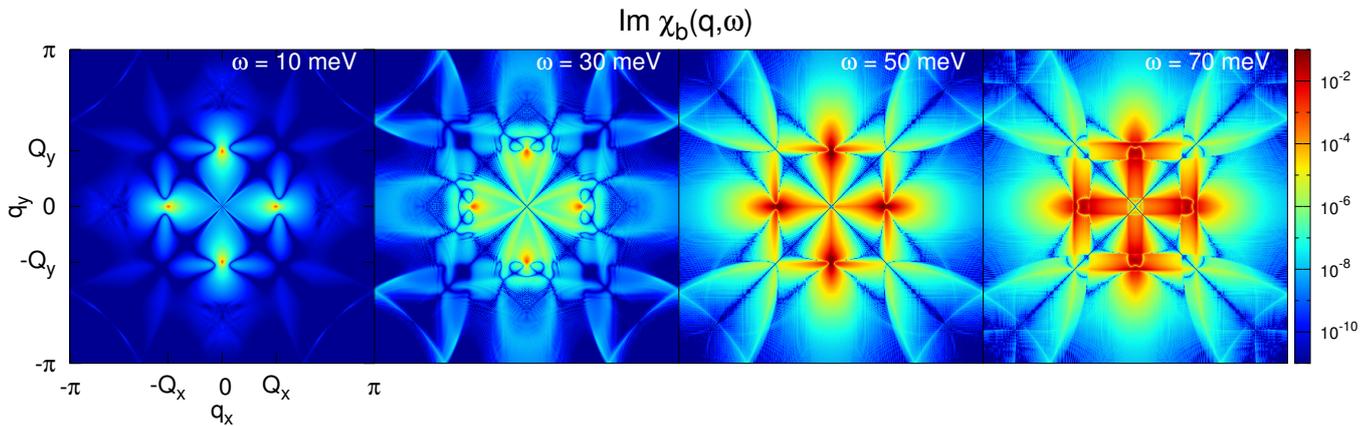}
\caption{Imaginary part of the contribution of the collective mode to the charge susceptibility $\chi_\eta$, plotted in the Brillouin zone for four different frequencies, with a very wide scale range. It peaks at the charge ordering wave vector $Q_0$, situated on the edge. Note that we are considering the same logarithmic scale for all four plots, as indicated on the right hand side. This shows that for a set frequency, the peak in $\chi_\eta$ is always situated close to $Q_0$, and that the strength of this peaks grows larger when crossing the Stoner continuum, at about 50 meV. Moreover, the texture of the continuum $B$ is clearly visible, and much smaller in magnitude.}
\label{fig:bubble-scale} 
\end{figure*}

\begin{figure*}
\centering
\includegraphics[width=18cm]{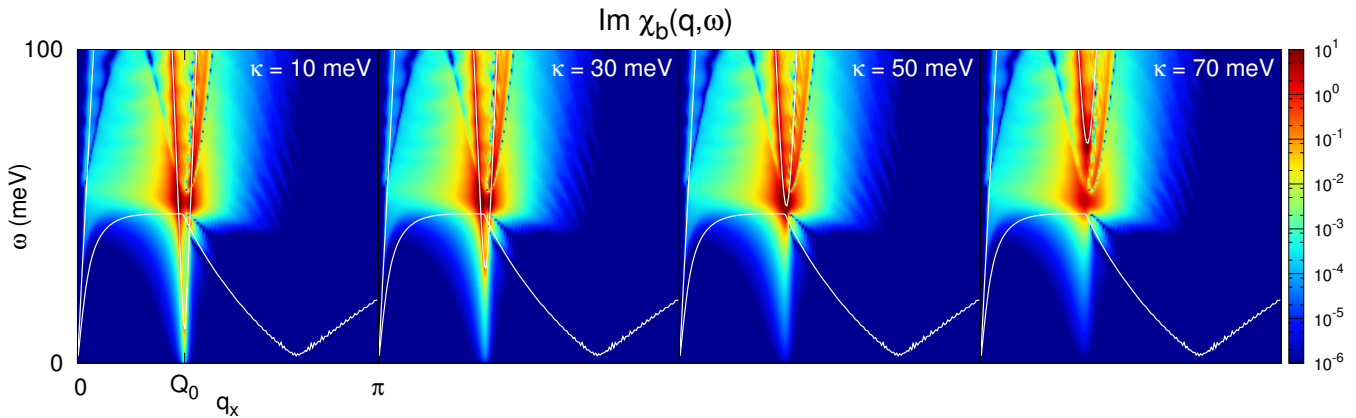}
\caption{Frequency dependence of the imaginary part of $\chi_\eta$ along the axis of the Brillouin zone, for a range of bosonic masses. We use the same logarithmic scale for all four plots, as indicated on the right hand side. In grey are also plotted the bottom and top of the Stoner continuum, along with the line where the bosonic propagator diverges. Note that for low masses, the response peaks at two spots just above the continuum bottom line. These two spots merge when the mass is raised. The structure of the continuum $B$, as well as the fennel of the bosonic mode, are visible and lower in magnitude.}
\label{fig:mass-evolution-scale} 
\end{figure*}

Here we display again the evolution of the contribution of the collective mode to the charge susceptibility $\chi_\eta$ in the Brillouin zone and along the edge with frequency dependence, but with a much wider scale (Fig. \ref{fig:bubble-scale} and \ref{fig:mass-evolution-scale}). This allows to see more clearly how the shape of $B$ influences $\chi_\eta$. However, such a scale also shows that there is residual weight below the Stoner continuum. This stems from the fact that we have used a residual scattering of 2 meV for plotting the propagators. This means in particular that any weight under this energy is not physical, and the weight seen under the Stoner continuum falls into this category. Figures \ref{fig:bubble-scale} and \ref{fig:mass-evolution-scale} therefore allow to make a connection between the figures and the mathematical expression for $\chi_\eta$ are not to be taken as actual response plots.

\section{Evolution of the continuum bubble with the size of the superconducting gap}
\label{Evolution of the continuum bubble with the size of the superconducting gap}

\begin{figure*}
\centering
\includegraphics[width=13.5cm]{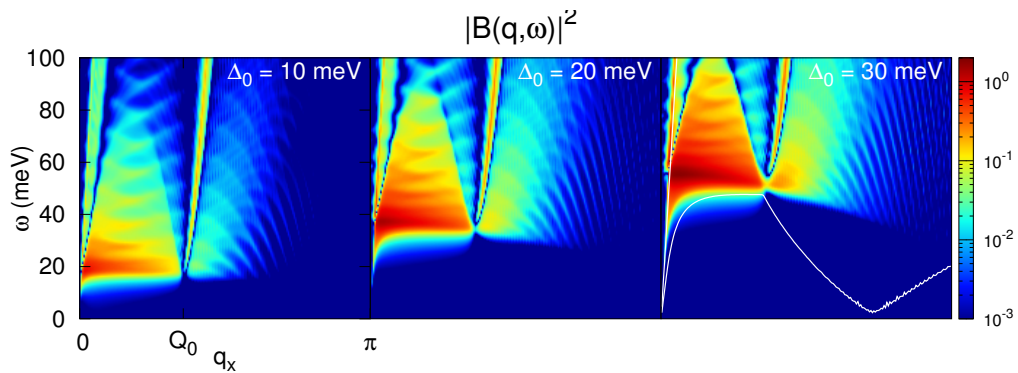}
\caption{Continuum bubble squared $|B|^2$ as a function of frequency along the axis of the Brillouin zone, plotted for various values of the superconducting gap. The white solid lines are the top and bottom of the Stoner continuum which for clarity we only show for $\Delta_0 = 30$ meV.}
\label{fig:bubble-delta} 
\end{figure*}

Here we show in more details the evolution of the continuum bubble when varying the superconducting gap. Fig.\ \ref{fig:bubble-delta} shows the frequency dependence of $|B|^2$ along the axis of the Brillouin zone. We see that, as expected from the fact that the superconducting gap is a coefficient of $B$, the magnitude of $|B|^2$ grows with $\Delta_0$. Moreover, as expected, the lower boundary of the Stoner continuum $\min_\textbf{k} \left( E_{-\textbf{k}+\textbf{q}} + E_\textbf{k} \right)$ also rises with $\Delta_0$. Something which would have been more difficult to directly see from the expression of $B$ is that the maximum weight of $B$ follows closely this boundary, and therefore that the peak in $\chi_\eta$ will also follow the evolution of $\Delta_0$.

\section{Decomposition of the Stoner continuum}
\label{Decomposition of the Stoner continuum}

\begin{figure}
\centering
\includegraphics[width=7cm]{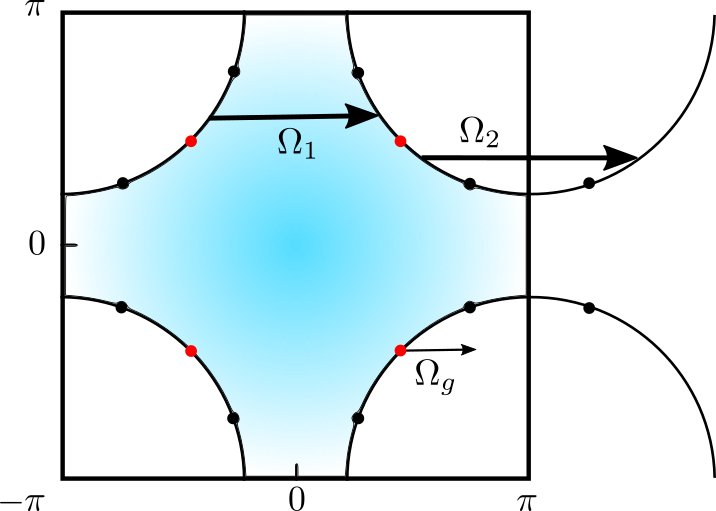}
\caption{Excitations corresponding to three frequency thresholds. The first two are related to excitations between points on the Fermi surface, and the third to excitations starting from the superconducting node, pictured in red.}
\label{fig:brillouin-zone} 
\end{figure}

The lower boundary of the Stoner continuum can be related to specific excitations between points on the Fermi surface. This was studied in particular close to the $(\pi,\pi)$ point of the Fermi surface, in relation to the neutron resonance measured at this point \cite{Morr2000, Eremin:2005ba}. More specifically, one can calculate the value of $E_{-\textbf{k}+\textbf{q}} + E_\textbf{k}$ for $\textbf{k}$ on the Fermi surface, which gives $|\Delta_{-\textbf{k}+\textbf{q}}| + |\Delta_\textbf{k}|$. We calculated this for two sets of wave-vectors $\textbf{q}$ parallel to the Brillouin zone axis, as displayed on Fig.\ \ref{fig:brillouin-zone}, and obtained the two corresponding frequencies $\Omega_1$ and $\Omega_2$.

\begin{figure}
\centering
\includegraphics[width=7cm]{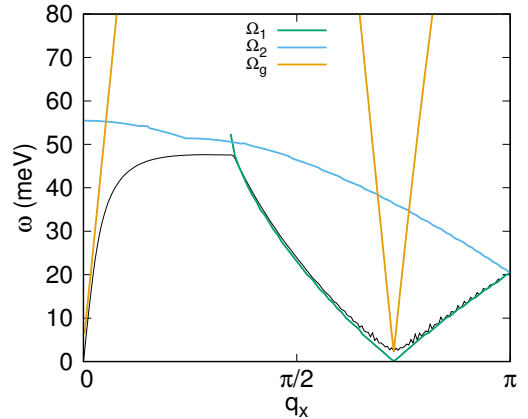}
\caption{Momentum dependence of three frequency thresholds on the axis of the Brillouin zone. The three thresholds correspond to the excitations displayed in Fig.\ \ref{fig:brillouin-zone}. The first two relate to excitations between two points on the Fermi surface, while the third one corresponds to excitations starting from the superconducting node. The black line is the bottom of the Stoner continuum.}
\label{fig:continuum-decomposition} 
\end{figure}

The result for $\Omega_1$ and $\Omega_2$ is shown on Fig.\ \ref{fig:continuum-decomposition}, along with the lower boundary of the Stoner continuum, and we can see that for high momenta the boundary of the continuum fits $\Omega_1$ very well. The small discrepancy close to the point where $\Omega_1$ softens is due to the discrete sampling of the Brillouin zone in the minimisation procedure. However at low momenta, neither $\Omega_1$ nor $\Omega_2$ follow the boundary of the continuum.This can be understood by simply looking at the shape of the Fermi surface displayed on Fig.\ \ref{fig:brillouin-zone}. Indeed, if we want to understand the low momentum limit of Fig.\ \ref{fig:continuum-decomposition}, we can consider asymptotically small vectors linking two points of the Fermi surface, which would therefore be tangent to it. Moreover we want only to consider vectors which are parallel to the Brillouin zone axis, since it is the plane on which we are studying the continuum boundary. The only wave-vectors that fit this description are the ones crossing the vertical edges of the Brillouin zone. There, the superconducting gap is large, and hence we cannot obtain the continuum line which goes to zero at $\textbf{q}=(0,0)$.

However it is possible to understand where this part of the continuum boundary comes from by considering a different limit. Instead of considering points on the Fermi surface, where $\xi_\textbf{k}=0$, we can consider points on the diagonal of the Brillouin zone, where the superconducting gap cancels: $\Delta_\textbf{k} = 0$. We therefore take $\textbf{k}$ at a superconducting node (Fig.\ \ref{fig:brillouin-zone}) and plot $\Omega_g = |\Delta_{-\textbf{k}+\textbf{q}}| + |\Delta_\textbf{k}|$ where $\textbf{q}$ is a vector along the axis of the Brillouin zone. The result is displayed on Fig.\ \ref{fig:continuum-decomposition}, and clearly shows that $\Omega_3$ closely follows the low-momentum limit of the continuum boundary.

It is clear on Fig.\ \ref{fig:continuum-decomposition} that $\Omega_2$ cannot be related to the shape of the lower boundary of the continuum. However we can clearly see its influence on $B$. Indeed, Fig.\ \ref{fig:continuum-energy} shows that, at high momentum, $|B|^2$ is very small above the continuum boundary, and only grows above a line which corresponds exactly to $\Omega_2$.

\begin{figure*}
\centering
\includegraphics[width=16cm]{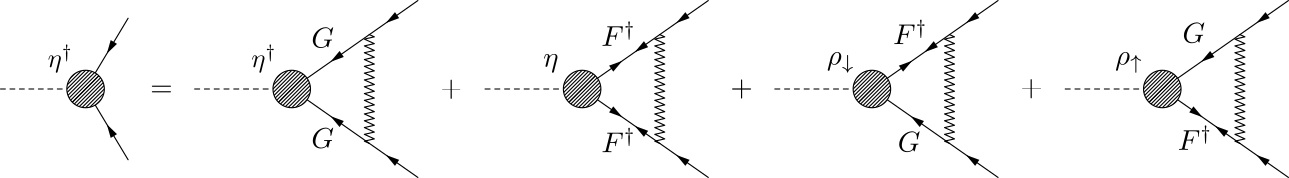}
\caption{Diagrams for the scattering in the particle-particle channel, corresponding to Eq.\ \eqref{eq:equation-eta-dagger}. The zigzag line represents the antiferromagnetic interaction.}
\label{fig:appendix-diagrams-eta-dagger} 
\end{figure*}

\begin{figure*}
\centering
\includegraphics[width=16cm]{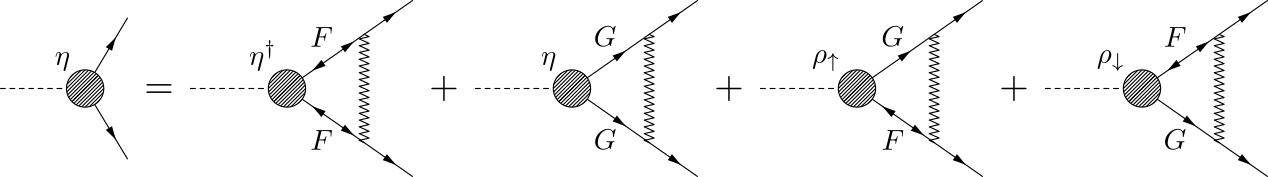}
\caption{Diagrams for the scattering in the hole-hole channel, corresponding to Eq.\ \eqref{eq:equation-eta}. The zigzag line represents the antiferromagnetic interaction.}
\label{fig:appendix-diagrams-eta} 
\end{figure*}

\begin{figure*}
\centering
\includegraphics[width=16cm]{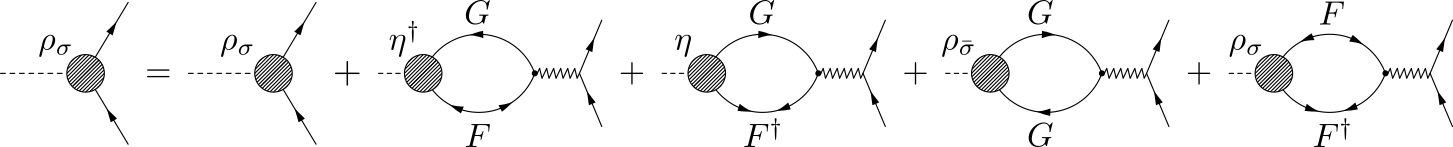}
\caption{Diagrams for the scattering in the particle-hole channel, corresponding to Eq.\ \eqref{eq:equation-rho}. The zigzag line represents the antiferromagnetic interaction.}
\label{fig:appendix-diagrams-rho} 
\end{figure*}

\bibliographystyle{longbibliography}
\bibliography{Cuprates}

\begin{thebibliography}{10}

\bibitem{Alloul89}
H.~Alloul, T.~Ohno, and P.~Mendels, ``$^{89}\mathrm{Y}$ nmr evidence for a
  fermi-liquid behavior in
  ${\mathrm{yba}}_{2}$${\mathrm{cu}}_{3}$${\mathrm{o}}_{6+\mathrm{x}}$,'' {\em
  Phys. Rev. Lett.}, vol.~63, pp.~1700--1703, Oct 1989.

\bibitem{Alloul91}
H.~Alloul, P.~Mendels, H.~Casalta, J.~F. Marucco, and J.~Arabski,
  ``Correlations between magnetic and superconducting properties of
  zn-substituted
  ${\mathrm{yba}}_{2}$${\mathrm{cu}}_{3}$${\mathrm{o}}_{6+\mathit{x}}$,'' {\em
  Phys. Rev. Lett.}, vol.~67, pp.~3140--3143, Nov 1991.

\bibitem{Warren89}
W.~W. Warren, R.~E. Walstedt, G.~F. Brennert, R.~J. Cava, R.~Tycko, R.~F. Bell,
  and G.~Dabbagh, ``Cu spin dynamics and superconducting precursor effects in
  planes above ${T}_{c}$ in
  ${\mathrm{yba}}_{2}$${\mathrm{cu}}_{3}$${\mathrm{o}}_{6.7}$,'' {\em Phys.
  Rev. Lett.}, vol.~62, pp.~1193--1196, Mar 1989.

\bibitem{Campuzano98}
J.~C. Campuzano, M.~R. Norman, H.~Ding, M.~Randeria, T.~Yokoya, T.~Takeuchi,
  T.~Takahashi, T.~Mochiku, K.~Kadowaki, P.~Guptasarma, and D.~G. Hinks,
  ``Destruction of the fermi surface in underdoped high-tc superconductors,''
  {\em Nature}, vol.~392, pp.~157--160, Mar. 1998.

\bibitem{Vishik:2012cc}
I.~M. Vishik, M.~Hashimoto, R.-H. He, W.-S. Lee, F.~Schmitt, D.~Lu, R.~G.
  Moore, C.~Zhang, W.~Meevasana, T.~Sasagawa, S.~Uchida, K.~Fujita, S.~Ishida,
  M.~Ishikado, Y.~Yoshida, H.~Eisaki, Z.~Hussain, T.~P. Devereaux, and Z.-X.
  Shen, ``{Phase competition in trisected superconducting dome},'' {\em Proc.
  Natl. Acad. Sci.}, vol.~109, pp.~18332--18337, Nov. 2012.

\bibitem{Vishik12}
I.~M. Vishik, M.~Hashimoto, R.-H. He, W.-S. Lee, F.~Schmitt, D.~Lu, R.~G.
  Moore, C.~Zhang, W.~Meevasana, T.~Sasagawa, S.~Uchida, K.~Fujita, S.~Ishida,
  M.~Ishikado, Y.~Yoshida, H.~Eisaki, Z.~Hussain, T.~P. Devereaux, and Z.-X.
  Shen, ``{Phase competition in trisected superconducting dome},'' {\em Proc.
  Natl. Acad. Sci.}, vol.~109, pp.~18332--18337, Nov. 2012.

\bibitem{Yoshida:2012kh}
T.~Yoshida, M.~Hashimoto, I.~M~Vishik, Z.-X. Shen, and A.~Fujimori,
  ``{Pseudogap, Superconducting Gap, and Fermi Arc in High-Tc Cuprates Revealed
  by Angle-Resolved Photoemission Spectroscopy},'' {\em J. Phys. Soc. Jpn.},
  vol.~81, p.~011006, Jan. 2012.

\bibitem{He14}
Y.~He, Y.~Yin, M.~Zech, A.~Soumyanarayanan, M.~M. Yee, T.~Williams, M.~C.
  Boyer, K.~Chatterjee, W.~D. Wise, I.~Zeljkovic, T.~Kondo, T.~Takeuchi,
  H.~Ikuta, P.~Mistark, R.~S. Markiewicz, A.~Bansil, S.~Sachdev, E.~W. Hudson,
  and J.~E. Hoffman, ``{Fermi Surface and Pseudogap Evolution in a Cuprate
  Superconductor},'' {\em Science}, vol.~344, no.~6184, pp.~608--611, 2014.

\bibitem{Vishik14}
I.~M. Vishik, N.~Bari\ifmmode \check{s}\else \v{s}\fi{}i\ifmmode~\acute{c}\else
  \'{c}\fi{}, M.~K. Chan, Y.~Li, D.~D. Xia, G.~Yu, X.~Zhao, W.~S. Lee,
  W.~Meevasana, T.~P. Devereaux, M.~Greven, and Z.-X. Shen, ``Angle-resolved
  photoemission spectroscopy study of
  $\mathrm{Hg}{\mathrm{ba}}_{2}\mathrm{Cu}{\mathrm{o}}_{4+\ensuremath{\delta}}$,''
  {\em Phys. Rev. B}, vol.~89, p.~195141, May 2014.

\bibitem{Doiron-Leyraud07}
N.~Doiron-Leyraud, C.~Proust, D.~LeBoeuf, J.~Levallois, J.-B. Bonnemaison,
  R.~Liang, D.~A. Bonn, W.~N. Hardy, and L.~Taillefer, ``{Quantum oscillations
  and the Fermi surface in an underdoped high-T$_c$ superconductor},'' {\em
  Nature}, vol.~447, pp.~565--568, May 2007.

\bibitem{LeBoeuf07}
D.~LeBoeuf, N.~Doiron-Leyraud, J.~Levallois, R.~Daou, J.~B. Bonnemaison, N.~E.
  Hussey, L.~Balicas, B.~J. Ramshaw, R.~Liang, D.~A. Bonn, W.~N. Hardy,
  S.~Adachi, C.~Proust, and L.~Taillefer, ``{Electron pockets in the Fermi
  surface of hole-doped high-T$_c$ superconductors},'' {\em Nature}, vol.~450,
  pp.~533--536, Nov. 2007.

\bibitem{LeBoeuf11}
D.~LeBoeuf, N.~Doiron-Leyraud, B.~Vignolle, M.~Sutherland, B.~J. Ramshaw,
  J.~Levallois, R.~Daou, F.~Lalibert\'e, O.~Cyr-Choini\`ere, J.~Chang, Y.~J.
  Jo, L.~Balicas, R.~Liang, D.~A. Bonn, W.~N. Hardy, C.~Proust, and
  L.~Taillefer, ``Lifshitz critical point in the cuprate superconductor
  yba2cu3oy from high-field hall effect measurements,'' {\em Phys. Rev. B},
  vol.~83, p.~054506, Feb 2011.

\bibitem{Laliberte11}
F.~Lalibert{\'e}, J.~Chang, N.~Doiron-Leyraud, E.~Hassinger, R.~Daou,
  M.~Rondeau, B.~J. Ramshaw, R.~Liang, D.~A. Bonn, W.~N. Hardy, S.~Pyon,
  T.~Takayama, H.~Takagi, I.~Sheikin, L.~Malone, C.~Proust, K.~Behnia, and
  L.~Taillefer, ``Fermi-surface reconstruction by stripe order in cuprate
  superconductors,'' {\em Nat. Commun.}, vol.~2, p.~432, 08 2011.

\bibitem{Sebastian12}
S.~E. Sebastian, N.~Harrison, R.~Liang, D.~A. Bonn, W.~N. Hardy, C.~H. Mielke,
  and G.~G. Lonzarich, ``Quantum oscillations from nodal bilayer magnetic
  breakdown in the underdoped high temperature superconductor
  ${\mathrm{yba}}_{2}{\mathrm{cu}}_{3}{\mathrm{o}}_{6+x}$,'' {\em Phys. Rev.
  Lett.}, vol.~108, p.~196403, May 2012.

\bibitem{Doiron-Leyraud13}
N.~Doiron-Leyraud, S.~Lepault, O.~Cyr-Choini\`ere, B.~Vignolle,
  G.~Grissonnanche, F.~Lalibert\'e, J.~Chang, N.~Bari\ifmmode \check{s}\else
  \v{s}\fi{}i\ifmmode~\acute{c}\else \'{c}\fi{}, M.~K. Chan, L.~Ji, X.~Zhao,
  Y.~Li, M.~Greven, C.~Proust, and L.~Taillefer, ``{Hall, Seebeck, and Nernst
  Coefficients of Underdoped
  ${\mathrm{HgBa}}_{2}{\mathrm{CuO}}_{4+\ensuremath{\delta}}$: Fermi-Surface
  Reconstruction in an Archetypal Cuprate Superconductor},'' {\em Phys. Rev.
  X}, vol.~3, p.~021019, Jun 2013.

\bibitem{Barisic2013}
N.~{Bari{\v s}i{\'c}}, S.~{Badoux}, M.~K. {Chan}, C.~{Dorow}, W.~{Tabis},
  B.~{Vignolle}, G.~{Yu}, J.~{B{\'e}ard}, X.~{Zhao}, C.~{Proust}, and
  M.~{Greven}, ``{Universal quantum oscillations in the underdoped cuprate
  superconductors},'' {\em Nature Physics}, vol.~9, pp.~761--764, Dec. 2013.

\bibitem{Grissonnanche:2015tl}
G.~{Grissonnanche}, F.~{Laliberte}, S.~{Dufour-Beausejour}, A.~{Riopel},
  S.~{Badoux}, M.~{Caouette-Mansour}, M.~{Matusiak}, A.~{Juneau-Fecteau},
  P.~{Bourgeois-Hope}, O.~{Cyr-Choiniere}, J.~C. {Baglo}, B.~J. {Ramshaw},
  R.~{Liang}, D.~A. {Bonn}, W.~N. {Hardy}, S.~{Kramer}, D.~{LeBoeuf},
  D.~{Graf}, N.~{Doiron-Leyraud}, and L.~{Taillefer}, ``{Onset field for
  Fermi-surface reconstruction in the cuprate superconductor YBCO},'' {\em
  ArXiv e-prints}, Aug. 2015.

\bibitem{Ghiringhelli12}
G.~Ghiringhelli, M.~Le~Tacon, M.~Minola, S.~Blanco-Canosa, C.~Mazzoli, N.~B.
  Brookes, G.~M. De~Luca, A.~Frano, D.~G. Hawthorn, F.~He, T.~Loew, M.~M. Sala,
  D.~C. Peets, M.~Salluzzo, E.~Schierle, R.~Sutarto, G.~A. Sawatzky,
  E.~Weschke, B.~Keimer, and L.~Braicovich, ``{Long-Range Incommensurate Charge
  Fluctuations in (Y,Nd)Ba$_2$Cu$_3$O$_{6+x}$},'' {\em Science}, vol.~337,
  no.~6096, pp.~821--825, 2012.

\bibitem{Chang12}
J.~Chang, E.~Blackburn, A.~T. Holmes, N.~B. Christensen, J.~Larsen, J.~Mesot,
  R.~Liang, D.~A. Bonn, W.~N. Hardy, A.~Watenphul, M.~v. Zimmermann, E.~M.
  Forgan, and S.~M. Hayden, ``{Direct observation of competition between
  superconductivity and charge density wave order in
  YBa$_2$Cu$_3$O$_{6.67}$},'' {\em Nat Phys}, vol.~8, pp.~871--876, 12 2012.

\bibitem{Achkar12}
A.~J. Achkar, R.~Sutarto, X.~Mao, F.~He, A.~Frano, S.~Blanco-Canosa,
  M.~Le~Tacon, G.~Ghiringhelli, L.~Braicovich, M.~Minola, M.~Moretti~Sala,
  C.~Mazzoli, R.~Liang, D.~A. Bonn, W.~N. Hardy, B.~Keimer, G.~A. Sawatzky, and
  D.~G. Hawthorn, ``Distinct charge orders in the planes and chains of
  ortho-iii-ordered
  ${\mathrm{yba}}_{2}{\mathrm{cu}}_{3}{\mathbf{o}}_{6\mathbf{+}\ensuremath{\delta}}$
  superconductors identified by resonant elastic x-ray scattering,'' {\em Phys.
  Rev. Lett.}, vol.~109, p.~167001, Oct 2012.

\bibitem{Blanco-Canosa13}
S.~Blanco-Canosa, A.~Frano, T.~Loew, Y.~Lu, J.~Porras, G.~Ghiringhelli,
  M.~Minola, C.~Mazzoli, L.~Braicovich, E.~Schierle, E.~Weschke, M.~Le~Tacon,
  and B.~Keimer, ``{Momentum-Dependent Charge Correlations in
  ${\mathrm{YBa}}_{2}{\mathrm{Cu}}_{3}{\mathrm{O}}_{6+\ensuremath{\delta}}$
  Superconductors Probed by Resonant X-Ray Scattering: Evidence for Three
  Competing Phases},'' {\em Phys. Rev. Lett.}, vol.~110, p.~187001, May 2013.

\bibitem{Blackburn13a}
E.~Blackburn, J.~Chang, M.~H\"ucker, A.~T. Holmes, N.~B. Christensen, R.~Liang,
  D.~A. Bonn, W.~N. Hardy, U.~R\"utt, O.~Gutowski, M.~v. Zimmermann, E.~M.
  Forgan, and S.~M. Hayden, ``X-ray diffraction observations of a
  charge-density-wave order in superconducting ortho-ii
  ${\mathrm{yba}}_{2}{\mathrm{cu}}_{3}{\mathbf{o}}_{6.54}$ single crystals in
  zero magnetic field,'' {\em Phys. Rev. Lett.}, vol.~110, p.~137004, Mar 2013.

\bibitem{Blackburn13b}
E.~Blackburn, J.~Chang, A.~H. Said, B.~M. Leu, R.~Liang, D.~A. Bonn, W.~N.
  Hardy, E.~M. Forgan, and S.~M. Hayden, ``Inelastic x-ray study of phonon
  broadening and charge-density wave formation in ortho-ii-ordered
  yba${}_{2}$cu${}_{3}$o${}_{6.54}$,'' {\em Phys. Rev. B}, vol.~88, p.~054506,
  Aug 2013.

\bibitem{Thampy13}
V.~Thampy, S.~Blanco-Canosa, M.~Garc\'{i}a-Fern\'andez, M.~P.~M. Dean, G.~D.
  Gu, M.~F\"orst, T.~Loew, B.~Keimer, M.~Le~Tacon, S.~B. Wilkins, and J.~P.
  Hill, ``{Comparison of charge modulations in
  La${}_{1.875}$Ba${}_{0.125}$CuO${}_{4}$ and
  YBa${}_{2}$Cu${}_{3}$O${}_{6.6}$},'' {\em Phys. Rev. B}, vol.~88, p.~024505,
  Jul 2013.

\bibitem{Blanco-Canosa14}
S.~Blanco-Canosa, A.~Frano, E.~Schierle, J.~Porras, T.~Loew, M.~Minola,
  M.~Bluschke, E.~Weschke, B.~Keimer, and M.~Le~Tacon, ``Resonant x-ray
  scattering study of charge-density wave correlations in
  ${\mathrm{yba}}_{2}{\mathrm{cu}}_{3}{\mathrm{o}}_{6+x}$,'' {\em Phys. Rev.
  B}, vol.~90, p.~054513, Aug 2014.

\bibitem{Tabis14}
W.~Tabis, Y.~Li, M.~{Le Tacon}, L.~Braicovich, A.~Kreyssig, M.~Minola,
  G.~Dellea, E.~Weschke, M.~J. Veit, M.~Ramazanoglu, A.~I. Goldman, T.~Schmitt,
  G.~Ghiringhelli, N.~{Bari{\v s}i{\'c}}, M.~K. Chan, C.~J. Dorow, G.~Yu,
  X.~Zhao, B.~Keimer, and M.~Greven, ``Charge order and its connection with
  fermi-liquid charge transport in a pristine high-tc cuprate,'' {\em Nat.
  Commun.}, vol.~5, p.~5875, 12 2014.

\bibitem{Comin14}
R.~Comin, A.~Frano, M.~M. Yee, Y.~Yoshida, H.~Eisaki, E.~Schierle, E.~Weschke,
  R.~Sutarto, F.~He, A.~Soumyanarayanan, Y.~He, M.~Le~Tacon, I.~S. Elfimov,
  J.~E. Hoffman, G.~A. Sawatzky, B.~Keimer, and A.~Damascelli, ``Charge order
  driven by fermi-arc instability in bi2sr2-xlaxcu06,'' {\em Science},
  vol.~343, no.~6169, pp.~390--392, 2014.

\bibitem{Comin14a}
R.~Comin, R.~Sutarto, F.~He, E.~H. da~Silva~Neto, L.~Chauviere, A.~Frano,
  R.~Liang, W.~N. Hardy, D.~A. Bonn, Y.~Yoshida, H.~Eisaki, A.~J. Achkar, D.~G.
  Hawthorn, B.~Keimer, G.~A. Sawatzky, and A.~Damascelli, ``Symmetry of charge
  order in cuprates,'' {\em Nat Mater}, vol.~14, pp.~796--800, 08 2015.

\bibitem{Comin:2015vc}
R.~Comin and A.~Damascelli, ``{Resonant x-ray scattering studies of charge
  order in cuprates},'' Sept. 2015.

\bibitem{Comin:2015ca}
R.~Comin, R.~Sutarto, F.~He, E.~H. da~Silva~Neto, L.~Chauviere, A.~Frano,
  R.~Liang, W.~N. Hardy, D.~A. Bonn, Y.~Yoshida, H.~Eisaki, A.~J. Achkar, D.~G.
  Hawthorn, B.~Keimer, G.~A. Sawatzky, and A.~Damascelli, ``{Symmetry of charge
  order in cuprates},'' {\em Nature Materials}, vol.~14, pp.~796--800, May
  2015.

\bibitem{Forgan2015}
E.~M. {Forgan}, E.~{Blackburn}, A.~T. {Holmes}, A.~{Briffa}, J.~{Chang},
  L.~{Bouchenoire}, S.~D. {Brown}, R.~{Liang}, D.~{Bonn}, W.~N. {Hardy}, N.~B.
  {Christensen}, M.~v. {Zimmermann}, M.~{Huecker}, and S.~M. {Hayden}, ``{The
  nature of the charge density waves in under-doped YBa$\_2$Cu$\_3$O$\_{6.54}$
  revealed by X-ray measurements of the ionic displacements},'' {\em ArXiv
  e-prints}, Apr. 2015.

\bibitem{Chang16}
J.~Chang, E.~Blackburn, O.~Ivashko, A.~T. Holmes, N.~B. Christensen, M.~Hucker,
  R.~Liang, D.~A. Bonn, W.~N. Hardy, U.~Rutt, M.~v. Zimmermann, E.~M. Forgan,
  and H.~S. M., ``Magnetic field controlled charge density wave coupling in
  underdoped yba2cu3o6+x,'' {\em Nat. Commun.}, vol.~7, p.~11494, 02 2016.

\bibitem{Efetov13}
K.~B. Efetov, H.~Meier, and C.~P\'epin, ``Pseudogap state near a quantum
  critical point,'' {\em Nat. Phys.}, vol.~9, pp.~442--446, 2013.

\bibitem{Kloss15}
T.~Kloss, X.~Montiel, and C.~P\'epin, ``{SU(2) symmetry in a realistic
  spin-fermion model for cuprate superconductors},'' {\em Phys. Rev. B},
  vol.~91, p.~205124, May 2015.

\bibitem{Kloss:2016hu}
T.~Kloss, X.~Montiel, V.~S. de~Carvalho, H.~Freire, and C.~P{\'e}pin, ``{Charge
  orders, magnetism and pairings in the cuprate superconductors},'' {\em
  Reports on Progress in Physics}, vol.~79, Aug. 2016.

\bibitem{Montiel16}
X.~Montiel, T.~Kloss, and C.~Pepin, ``Effective su(2) theory for the pseudogap
  state,'' {\em Phys. Rev. B}, vol.~95, p.~104510, 2017.

\bibitem{Montiel2017sr}
X.~Montiel, T.~Kloss, and C.~P{\'e}pin, ``Local particle-hole pair excitations
  by su(2) symmetry fluctuations,'' {\em Scientific Reports}, vol.~7, p.~3477,
  2017.

\bibitem{Montiel:2016it}
X.~Montiel, T.~Kloss, and C.~P{\'e}pin, ``{Angle resolved photo-emission
  spectroscopy signature of the resonant excitonic state},'' {\em EPL
  (Europhysics Letters)}, vol.~115, p.~57001, Oct. 2016.

\bibitem{Montiel15a}
X.~Montiel, T.~Kloss, C.~P\'epin, S.~Benhabib, Y.~Gallais, and A.~Sacuto,
  ``{Raman scattering and SU(2) collective resonance in cuprate
  superconductors},'' {\em arXiv:1504.03951 [cond-mat.supr-con]}, 2015.

\bibitem{MontielNeutrons2017}
X.~Montiel and C.~P\'epin, ``Model for the neutron resonance in
  ${\mathrm{hgba}}_{2}{\mathrm{cuo}}_{4+\ensuremath{\delta}}$,'' {\em Phys.
  Rev. B}, vol.~96, p.~094529, Sep 2017.

\bibitem{Morice2017transport}
C.~Morice, X.~Montiel, and C.~P\'epin, ``Evolution of hall resistivity and
  spectral function with doping in the su(2) theory of cuprates,'' {\em Phys.
  Rev. B}, vol.~96, p.~134511, Oct 2017.

\bibitem{Chakraborty2018}
D.~{Chakraborty}, C.~{Morice}, and C.~{P{\'e}pin}, ``{The phase diagram of the
  underdoped cuprates at high magnetic field},'' {\em ArXiv e-prints}, Feb.
  2018.

\bibitem{Abanov00}
A.~Abanov and A.~V. Chubukov, ``{Spin-Fermion Model near the Quantum Critical
  Point: One-Loop Renormalization Group Results},'' {\em Phys. Rev. Lett.},
  vol.~84, pp.~5608--5611, Jun 2000.

\bibitem{Tchernyshyov2001}
O.~Tchernyshyov, M.~R. Norman, and A.~V. Chubukov, ``Neutron resonance in
  high-${T}_{c}$ superconductors is not the $\pi${} particle,'' {\em Phys. Rev.
  B}, vol.~63, p.~144507, Mar 2001.

\bibitem{Abanov03}
A.~Abanov, A.~V. Chubukov, and J.~Schmalian, ``{Quantum-critical theory of the
  spin-fermion model and its application to cuprates: Normal state analysis},''
  {\em Adv. Phys.}, vol.~52, no.~3, pp.~119--218, 2003.

\bibitem{Hao2009}
Z.~Hao and A.~V. Chubukov, ``{Resonance peak in neutron scattering experiments
  on the cuprates revisited: The case of exciton versus $\pi$ -resonance and
  magnetic plasmon},'' {\em Physical Review B - Condensed Matter and Materials
  Physics}, vol.~79, no.~22, pp.~1--5, 2009.

\bibitem{Metlitski10a}
M.~A. Metlitski and S.~Sachdev, ``{Quantum phase transitions of metals in two
  spatial dimensions. I. Ising-nematic order},'' {\em Phys. Rev. B}, vol.~82,
  p.~075127, Aug 2010.

\bibitem{Metlitski10b}
M.~A. Metlitski and S.~Sachdev, ``{Quantum phase transitions of metals in two
  spatial dimensions. II. Spin density wave order},'' {\em Phys. Rev. B},
  vol.~82, p.~075128, Aug 2010.

\bibitem{Lee06}
P.~A. Lee, N.~Nagaosa, and X.-G. Wen, ``{Doping a Mott insulator: Physics of
  high-temperature superconductivity},'' {\em Rev. Mod. Phys.}, vol.~78,
  pp.~17--85, Jan 2006.

\bibitem{Demler95}
E.~Demler and S.-C. Zhang, ``Theory of the resonant neutron scattering of
  high-${T}_{c}$ superconductors,'' {\em Phys. Rev. Lett.}, vol.~75,
  pp.~4126--4129, Nov 1995.

\bibitem{Demler1997}
E.~Demler, H.~Kohno, and S.-C. Zhang, ``{Pi excitation of the t-J model},''
  {\em Phys. Rev. B}, vol.~58, no.~9, pp.~1--33, 1997.

\bibitem{Zhang1997}
S.-C. Zhang, ``A unified theory based on so(5) symmetry of superconductivity
  and antiferromagnetism,'' {\em Science}, vol.~275, no.~5303, pp.~1089--1096,
  1997.

\bibitem{Zaleski2000}
T.~A. Zaleski and T.~K. Kope{\'{c}}, ``{Phase diagrams in the SO(5) quantum
  rotor theory of high-Tc superconductivity},'' {\em Physical Review B},
  vol.~62, no.~13, pp.~9059--9076, 2000.

\bibitem{Zaleski2001}
T.~Zaleski and T.~Kope{\'{c}}, ``{Magnetic correlation functions in SO(5)
  theory of high-Tc superconductivity},'' {\em Physical Review B}, vol.~64,
  no.~14, p.~144522, 2001.

\bibitem{Hu2001}
J.-P. Hu and S.-C. Zhang {\em Physical Review B}, vol.~64, no.~10, p.~100502,
  2001.

\bibitem{Demler2004}
E.~Demler, W.~Hanke, and S.-C. Zhang, ``{SO(5) theory of antiferromagnetism and
  superconductivity},'' {\em Rev. Mod. Phys.}, vol.~76, no.~3, pp.~909--974,
  2004.

\bibitem{Kivelson:2002er}
S.~A. Kivelson, D.-H. Lee, E.~Fradkin, and V.~Oganesyan, ``{Competing order in
  the mixed state of high-temperature superconductors},'' {\em Phys. Rev. B},
  vol.~66, p.~144516, Oct. 2002.

\bibitem{Zhang:2002hz}
Y.~Zhang, E.~Demler, and S.~Sachdev, ``{Competing orders in a magnetic field:?
  Spin and charge order in the cuprate superconductors},'' {\em Phys.l Rev. B},
  vol.~66, p.~094501, Sept. 2002.

\bibitem{Hayward14}
L.~E. Hayward, D.~G. Hawthorn, R.~G. Melko, and S.~Sachdev, ``{Angular
  Fluctuations of a Multicomponent Order Describe the Pseudogap of
  YBa$_2$Cu$_3$O$_{6+x}$},'' {\em Science}, vol.~343, no.~6177, pp.~1336--1339,
  2014.

\bibitem{PhysRevB.92.224504}
Y.~Caplan, G.~Wachtel, and D.~Orgad, ``Long-range order and pinning of
  charge-density waves in competition with superconductivity,'' {\em Phys. Rev.
  B}, vol.~92, p.~224504, Dec 2015.

\bibitem{PhysRevLett.119.107002}
Y.~Caplan and D.~Orgad, ``Dimensional crossover of charge-density wave
  correlations in the cuprates,'' {\em Phys. Rev. Lett.}, vol.~119, p.~107002,
  Sep 2017.

\bibitem{Vig2017}
S.~Vig, A.~Kogar, M.~Mitrano, A.~A. Husain, V.~Mishra, M.~S. Rak, L.~Venema,
  P.~D. Johnson, G.~D. Gu, E.~Fradkin, M.~R. Norman, and P.~Abbamonte,
  ``{Measurement of the dynamic charge response of materials using low-energy,
  momentum-resolved electron energy-loss spectroscopy (M-EELS)},'' {\em SciPost
  Phys.}, vol.~3, p.~026, 2017.

\bibitem{Mitrano2017}
M.~{Mitrano}, A.~A. {Husain}, S.~{Vig}, A.~{Kogar}, M.~S. {Rak}, S.~I.
  {Rubeck}, J.~{Schmalian}, B.~{Uchoa}, J.~{Schneeloch}, R.~{Zhong}, G.~D.
  {Gu}, and P.~{Abbamonte}, ``{Singular density fluctuations in the strange
  metal phase of a copper-oxide superconductor},'' Aug. 2017.

\bibitem{Moor2017}
A.~Moor, A.~F. Volkov, and K.~B. Efetov, ``Amplitude higgs mode and admittance
  in superconductors with a moving condensate,'' {\em Phys. Rev. Lett.},
  vol.~118, p.~047001, Jan 2017.

\bibitem{Montiel2017}
X.~Montiel, T.~Kloss, and C.~P\'epin, ``Effective su(2) theory for the
  pseudogap state,'' {\em Phys. Rev. B}, vol.~95, p.~104510, Mar 2017.

\bibitem{Vishik2012}
I.~M. Vishik, M.~Hashimoto, R.-H. He, W.-S. Lee, F.~Schmitt, D.~Lu, R.~G.
  Moore, C.~Zhang, W.~Meevasana, T.~Sasagawa, S.~Uchida, K.~Fujita, S.~Ishida,
  M.~Ishikado, Y.~Yoshida, H.~Eisaki, Z.~Hussain, T.~P. Devereaux, and Z.-X.
  Shen, ``{Phase competition in trisected superconducting dome},'' {\em Proc.
  Natl. Acad. Sci.}, vol.~109, pp.~18332--18337, 2012.

\bibitem{Dai2018}
Z.~Dai, Y.-H. Zhang, T.~Senthil, and P.~A. Lee, ``Pair-density waves,
  charge-density waves, and vortices in high-${T}_{c}$ cuprates,'' {\em Phys.
  Rev. B}, vol.~97, p.~174511, May 2018.

\bibitem{Sachdev98}
S.~Sachdev, {\em {Quantum Phase Transitions}}.
\newblock New York: Cambridge University Press, 1998.

\bibitem{Anderson:1963vi}
P.~W. Anderson, ``{Plasmons, gauge invariance, and mass},'' {\em Phys. Rev.},
  vol.~130, no.~1, pp.~439--442, 1963.

\bibitem{Meier13}
H.~Meier, M.~Einenkel, C.~P\'epin, and K.~B. Efetov, ``Effect of magnetic field
  on the competition between superconductivity and charge order below the
  pseudogap state,'' {\em Phys. Rev. B}, vol.~88, p.~020506, Jul 2013.

\bibitem{Demler1996}
E.~Demler, S.-C. Zhang, N.~Bulut, and D.~J. Scalapino, ``A new class of
  collective excitations of the hubbard model: η excitation of the negative-u
  model,'' {\em International Journal of Modern Physics B}, vol.~10, no.~17,
  pp.~2137--2166, 1996.

\bibitem{Schrieffer64}
J.~R. Schrieffer, {\em Theory of Superconductivity}.
\newblock Benjamin Reading, MA, 1964.

\bibitem{Norman07}
M.~R. Norman, ``Linear response theory and the universal nature of the magnetic
  excitation spectrum of the cuprates,'' {\em Phys. Rev. B}, vol.~75,
  p.~184514, May 2007.

\bibitem{Norman:1995dd}
M.~R. Norman, M.~Randeria, H.~Ding, and J.~C. Campuzano, ``{ Phenomenological
  models for the gap anisotropy of Bi 2 Sr 2 CaCu 2 O 8 as measured by
  angle-resolved photoemission spectroscopy },'' {\em Phys. Rev. B}, vol.~52,
  no.~1, pp.~615--622, 1995.

\bibitem{Doiron-Leyraud2017}
N.~Doiron-Leyraud, O.~Cyr-Choinière, S.~Badoux, A.~Ataei, C.~Collignon,
  A.~Gourgout, S.~Dufour-Beauséjour, F.~F. Tafti, F.~Laliberté, M.-E.
  Boulanger, M.~Matusiak, D.~Graf, M.~Kim, J.-S. Zhou, N.~Momono, T.~Kurosawa,
  H.~Takagi, and L.~Taillefer, ``Pseudogap phase of cuprate superconductors
  confined by fermi surface topology,'' {\em Nat. Comm.}, vol.~8, p.~2044,
  2017.

\bibitem{Bruus2004book}
H.~Bruus, K.~Flensberg, and O.~U. Press, {\em Many-Body Quantum Theory in
  Condensed Matter Physics: An Introduction}.
\newblock Oxford Graduate Texts, OUP Oxford, 2004.

\bibitem{Chaix2017}
L.~Chaix, G.~Ghiringhelli, Y.~Y. Peng, M.~Hashimoto, B.~Moritz, K.~Kummer,
  N.~B. Brookes, Y.~He, S.~Chen, S.~Ishida, Y.~Yoshida, H.~Eisaki, M.~Salluzzo,
  L.~Braicovich, Z.~X. Shen, T.~P. Devereaux, and W.~S. Lee, ``{Dispersive
  charge density wave excitations in Bi2Sr2CaCu2O8+$\delta$},'' {\em Nature
  Physics}, vol.~13, no.~10, pp.~952--956, 2017.

\bibitem{Zhou2018}
K.~Zhou, ``{In preparation},'' 2018.

\bibitem{Borisenko2004}
S.~V. Borisenko, A.~A. Kordyuk, A.~Koitzsch, M.~Knupfer, J.~Fink, H.~Berger,
  and C.~T. Lin, ``Time-reversal symmetry breaking?,'' {\em Nature}, vol.~431,
  2004.

\bibitem{Devereaux2016}
T.~P. Devereaux, A.~M. Shvaika, K.~Wu, K.~Wohlfeld, C.~J. Jia, Y.~Wang,
  B.~Moritz, L.~Chaix, W.-S. Lee, Z.-X. Shen, G.~Ghiringhelli, and
  L.~Braicovich, ``Directly characterizing the relative strength and momentum
  dependence of electron-phonon coupling using resonant inelastic x-ray
  scattering,'' {\em Phys. Rev. X}, vol.~6, p.~041019, Oct 2016.

\bibitem{Mitrano2018}
M.~{Mitrano}, S.~{Lee}, A.~A. {Husain}, L.~{Delacretaz}, M.~{Zhu}, G.~{de la
  Pe{\~n}a Munoz}, S.~{Sun}, Y.~I. {Joe}, A.~H. {Reid}, S.~F. {Wandel},
  G.~{Coslovich}, W.~{Schlotter}, T.~{van Driel}, J.~{Schneeloch}, G.~D. {Gu},
  S.~{Hartnoll}, N.~{Goldenfeld}, and P.~{Abbamonte}, ``{Ultrafast
  time-resolved x-ray scattering reveals diffusive charge order dynamics in
  La$\_{2-x}$Ba$\_x$CuO$\_4$},'' {\em ArXiv e-prints}, Aug. 2018.

\bibitem{Cooper1988}
S.~L. Cooper, F.~Slakey, M.~V. Klein, J.~P. Rice, E.~D. Bukowski, and D.~M.
  Ginsberg, ``Gap anisotropy and phonon self-energy effects in single-crystal
  $\mathrm{Y}{\mathrm{ba}}_{2}{\mathrm{cu}}_{3}{\mathrm{o}}_{7\ensuremath{-}\ensuremath{\delta}}$,''
  {\em Phys. Rev. B}, vol.~38, pp.~11934--11937, Dec 1988.

\bibitem{Staufer1992}
T.~Staufer, R.~Nemetschek, R.~Hackl, P.~M\"uller, and H.~Veith, ``Investigation
  of the superconducting order parameter in
  ${\mathrm{bi}}_{2}$${\mathrm{sr}}_{2}$${\mathrm{cacu}}_{2}$${\mathrm{o}}_{8}$
  single crystals,'' {\em Phys. Rev. Lett.}, vol.~68, pp.~1069--1072, Feb 1992.

\bibitem{Sacuto1997}
A.~Sacuto, R.~Combescot, N.~Bontemps, P.~Monod, V.~Viallet, and D.~Colson,
  ``Nodes of the superconducting gap probed by electronic raman scattering in
  hgba 2 cacu 2 o 6 + δ single crystals,'' {\em EPL (Europhysics Letters)},
  vol.~39, no.~2, p.~207, 1997.

\bibitem{Gallais04}
Y.~{Gallais}, A.~{Sacuto}, and D.~{Colson}, ``{Resonant Raman scattering in
  mercurate single crystals},'' {\em Physica C}, vol.~{408 - 410}, no.~0,
  pp.~785 -- 788, 2004.

\bibitem{LeTacon05b}
M.~Le~Tacon, A.~Sacuto, and D.~Colson, ``Two distinct electronic contributions
  in the fully symmetric raman response of high-$t_{c}$ cuprates,'' {\em Phys.
  Rev. B}, vol.~71, p.~100504, Mar 2005.

\bibitem{Morr2000}
D.~K. Morr and D.~Pines, ``Magnetic coherence as a universal feature of cuprate
  superconductors,'' {\em Phys. Rev. B}, vol.~62, pp.~15177--15182, Dec 2000.

\bibitem{Eremin:2005ba}
I.~Eremin, D.~Morr, A.~Chubukov, K.~Bennemann, and M.~Norman, ``{Novel neutron
  resonance mode in dx2-y2-wave superconductors},'' {\em Phys. Rev. Lett.},
  vol.~94, no.~14, p.~147001, 2005.

\end{thebibliography}

\end{document}